\documentclass[%
 reprint,
 superscriptaddress,
 amsmath,amssymb,
 aps,
]{revtex4-2}

\usepackage{graphicx}
\usepackage{xfrac}
\usepackage{hyperref}
\usepackage{braket}
\usepackage{booktabs}
\hypersetup{colorlinks,urlcolor=blue, linkcolor=blue}
 
\begin{document}


\title{Host-guest Crystal Engineering Tailors the Room Temperature Spin Dynamics in Molecular Quantum Devices}

\author{Ziqiu Huang}
\affiliation{Department of Materials and London Centre for Nanotechnology, Imperial College London, Prince Consort Road, London, SW7 2AZ, UK}

\author{Angus Cowley-Semple}
\affiliation{James Watt School of Engineering, University of Glasgow, Glasgow, G12 8QQ, UK.}

\author{Irena Nevjestic}
\affiliation{Department of Materials and London Centre for Nanotechnology, Imperial College London, Prince Consort Road, London, SW7 2AZ, UK}

\author{Yifan Yu}
\affiliation{Department of Materials and London Centre for Nanotechnology, Imperial College London, Prince Consort Road, London, SW7 2AZ, UK}

\author{Mark Oxborrow}
\email{m.oxborrow@imperial.ac.uk}
\affiliation{Department of Materials and London Centre for Nanotechnology, Imperial College London, Prince Consort Road, London, SW7 2AZ, UK}

\author{Sam L. Bayliss}
\email{sam.bayliss@glasgow.ac.uk}
\affiliation{James Watt School of Engineering, University of Glasgow, Glasgow, G12 8QQ, UK.}

\author{Sarah K. Mann}
\email{sarah.mann@glasgow.ac.uk}
\affiliation{James Watt School of Engineering, University of Glasgow, Glasgow, G12 8QQ, UK.}

\author{Max Attwood}
\email{m.attwood@imperial.ac.uk}
\affiliation{Department of Materials and London Centre for Nanotechnology, Imperial College London, Prince Consort Road, London, SW7 2AZ, UK}


\begin{abstract}
\section*{Abstract}

Molecular materials that enable coherent control over an electron's spin state at room temperature are promising candidates for quantum technologies, including quantum sensors and ultra-low noise microwave amplifiers, known as masers. Host–guest molecular crystals enable independent control of spin-active guests and their local environments to enhance molecular spin properties and so improve device performance. Using electron paramagnetic resonance and optically-detected magnetic resonance, we demonstrate the ability to tune triplet (de)population and spin-lattice relaxation by modulating host-dependent lattice rigidity and vibrational coupling to significantly reduce the operating requirements for building useful masers. Importantly, the most rigid host, picene, reveals the ability to slow spin–lattice relaxation without lengthening triplet lifetime, though at the cost of strain-induced line width broadening and reduced triplet spin polarisation. We also find that deuteration reduces the triplet resonance line width and vibrationally-mediated triplet depopulation. Consequently, we find that perdeuterated pentacene in perdeuterated \textit{para}-terphenyl is the most viable candidate for building a continuous wave maser. This work demonstrates host-guest engineering as an important and practical method for tuning the spin-dependent performance of room-temperature molecular quantum technologies.
\end{abstract}

\maketitle
\section{Introduction}

 Materials bearing unpaired electron spins are attractive building blocks for quantum sensing technologies due to their ability to couple to weak electromagnetic fields \cite{Degen2017, Yu2021, Aslam2023, Bonizzoni2024, Roberts2025}. To align with many real-world applications, sensing platforms require abundant and coherent spin states under ambient conditions to amplify their response far above background noise. This requirement is central to building masers, which are an established type of ultra-low-noise quantum-enhanced microwave amplifier that have been successfully demonstrated using solid-state spins \cite{RobertCClauss2008, oxborrow2012room, breeze2018continuous, AESiegman1964}. In the past decade or so, masers have been realised at room temperature by employing optically induced spin-selective intersystem crossing (ISC) to generate spin polarized triplet states in materials such as negatively-charged nitrogen vacancy (NV) diamond \cite{breeze2018continuous, Sherman2022}, silicon carbide \cite{Gottscholl2022, Gottscholl2023}, and pentacene-doped \textit{para}-terphenyl \cite{oxborrow2012room}. However, at room temperature, high gain and low-noise (i.e., useful) masing is yet to be demonstrated beyond a few milliseconds \cite{wu2020room}. This is because to build a useful maser, the gain medium must contain a high concentration of strongly polarized spins, which is difficult to maintain over longer periods of stimulated emission. Moreover, high spin concentrations give rise to spin-spin interactions that induce spin-dephasing and broaden the material's inhomogeneous emission line width beyond the bandwidth of the cavity, resulting in diminishing returns with increasing concentration. Therefore, low-noise continuous wave (CW) masing has only been achieved using parts-per-million-concentration NV centres in diamond with external cooling to reduce spin relaxation and spin dephasing \cite{Sherman2022, Day2024}, presenting a challenge for room-temperature applications. On the other hand, organic molecular materials such as pentacene-doped \textit{para}-terphenyl (Pc:PTP) can easily be synthesised at parts-per-thousand (i.e., 0.01-0.1\% mol/mol) concentrations and have enabled room-temperature, though pulsed, masers. Their CW operation has remained elusive due to long-lived unpolarised triplet spin-states that accumulate during stimulated emission. To compensate, high optical pump energies are required which can lead to chemical instability. Therefore, to realise useful CW masing, it is essential to improve the spin dynamics of this molecular platform to reduce unpolarised spin states and inhomogeneous broadening.

\begin{figure}
    \centering
    \includegraphics[width=\columnwidth]{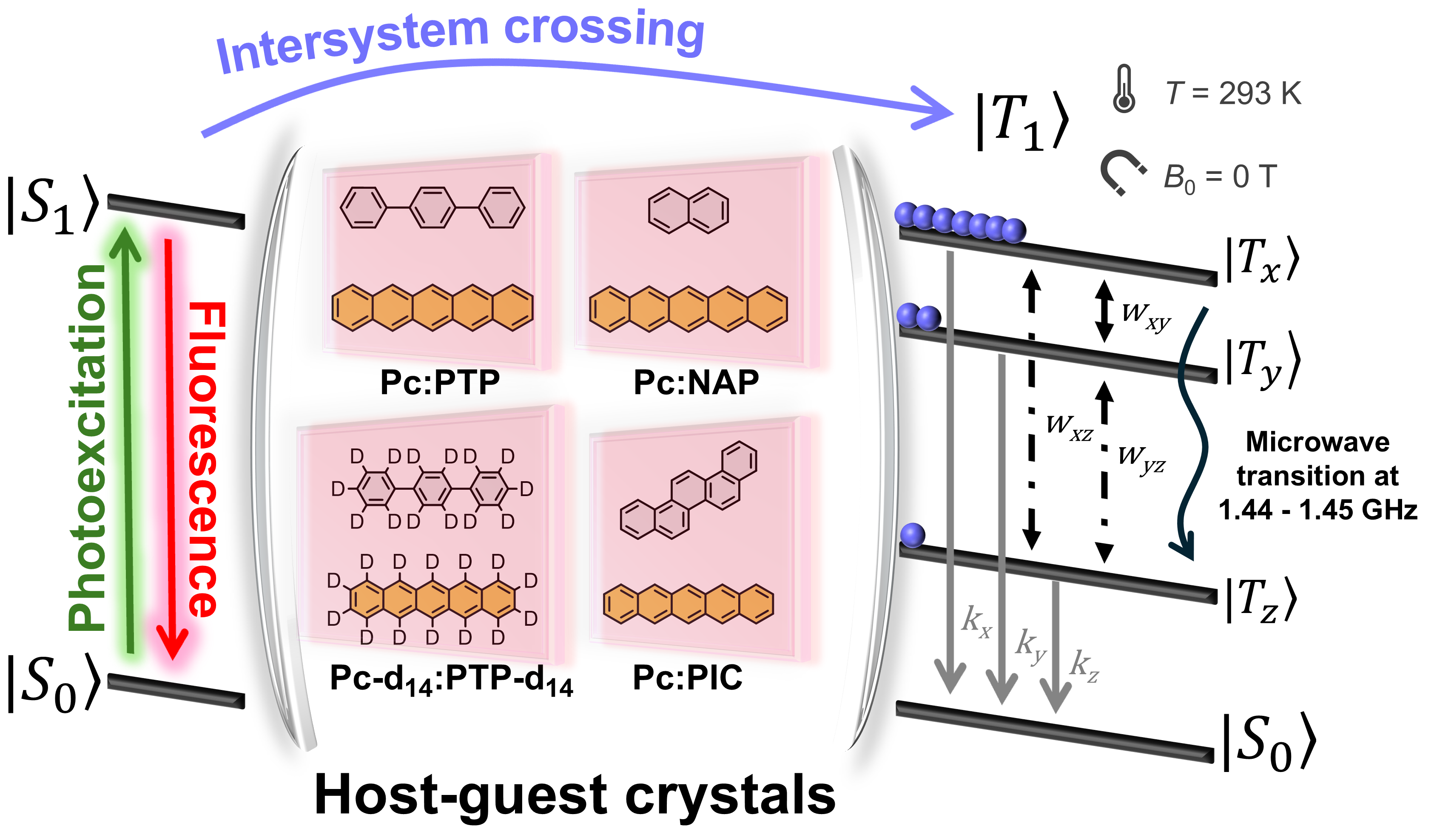}
    \caption{\textbf{Schematic of the photophysical processes in pentacene host–guest crystals.} A strongly spin polarised triplet state is formed after optical excitation from the ground ($\ket{S_0}$) to the excited ($\ket{S_1}$) singlet state, followed by spin-selective intersystem crossing (ISC) and relaxation to $\ket{T_1}$ \cite{sloop1981electron}. In $\ket{T_1}$, electron dipole interactions result in a zero-field splitting (ZFS), characterised by axial ($D \approx$ 1.45 GHz) and rhombic ($\lvert \textit{E} \rvert \approx$ 50 MHz) energies that split the triplet state into well-defined $\ket{T_x}$, $\ket{T_y}$ and $\ket{T_z}$ spin-sublevels. Spin-lattice relaxation between these sublevels ($\omega_{ij}$, where i and j are initial and final states) takes several 10s of microseconds, allowing the population inversion to persist long enough for resonant microwave photons to drive stimulated emission and thereby achieve microwave amplification. Here, we investigate the 1.45 GHz transition corresponding to the $\ket{T_x}$ $\leftrightarrow$ $\ket{T_z}$ transition at zero-applied magnetic field and room temperature.}
    \label{Jablonski}
    \vspace{2mm} 
    \hrule
    \vspace{1mm}
\end{figure}

To engineer a useful maser, the objective is to generate materials that minimise deactivation of the excited singlet state, $\ket{S_1}$, by fluorescence and non-radiative internal conversion (IC), and maximise ISC to populate either $\ket{T_x}$ or $\ket{T_y}$ triplet sublevels (see Figure \ref{Jablonski}). These maser-relevant states should also exhibit a narrow line width, and finally, triplet states that do not contribute to masing (e.g., $\ket{T_y}$ for the $\ket{T_x}\leftrightarrow\ket{T_z}$ transition in pentacene) should have relatively fast decay rates ($k_i$) so they can be recycled through the singlet state (i.e., ideally, $k_x < k_{y/z}$). In pentacene, the rate of ISC is known to depend, firstly, on the near-degeneracy of $\ket{S_1}$ and $\ket{T_2}$-states according to Fermi's Golden Rule, and secondly, on asymmetric and out-of-plane (OOP) vibrations, often termed ``butterfly'' modes \cite{Kohler1996, kryschi1992vibronically}. These vibrations reduce the local symmetry of pentacene's $\pi$-orbital system and mix $\sigma$-orbital character into the lowest energy excited states, such as $\ket{S_1}$. The net result is that otherwise symmetry-forbidden coupling between $\ket{S_1}$ and $\ket{T_2}$ is relaxed, enabling spin-selective ISC primarily to the $\ket{T_x}$ and $\ket{T_y}$ spin-sublevels. In contrast, $\ket{T_z}$ couples much less efficiently to $\pi$$\pi^*$-states through in-plane vibrations and C-H bending resulting in a smaller relative $\ket{T_z}$ population \cite{Clarke1976}. ISC from $\ket{T_{x/y/z}}$ to $\ket{S_0}$ is governed by identical vibrational considerations; hence, $k_z > k_y > k_x$ \cite{Wu2019UnravelingField, mann2025chemically}.

An established approach for modifying molecular spin dynamics is host-guest engineering, where spin-active guest molecules are controllably diluted into an inert host crystal that provides atmospheric protection and 3-dimensional orientational order. Pentacene has been doped into PTP \cite{patterson1984intersystem}, naphthalene \cite{VanStrien1980, Lambert1980, Kummer1996, takeda2002zero, Hautle2026}, anthracene \cite{Wakayama1973, Wakayama1974}, tetracene \cite{burgos1977heterofission, Zeiser2020, Wagner2024, wagner2024readout}, pyrene \cite{Kawaoka1964}, 1,3,5-tri(1-naphthyl)benzene \cite{schroder2022pentacene}, benzoic acid \cite{corval1992deuteration, Ong1993TemperatureCrystals, Ong1994}, \textit{trans}-1,4-distyrylbenzene \cite{wang2009doped} and more recently, picene \cite{moro2022room}, dibenz[a,h]anthracene \cite{Tateishi2026} and hexacene \cite{Unger2024}. Generally, hosts that do not promote charge dissociation and facilitate OOP vibrations and symmetry-breaking for pentacene dopants, such as PTP, promote ISC whilst others such as naphthalene and benzoic acid promote fluorescence \cite{Takeda2002, kryschi1992vibronically, Astilean1994}; though in all cases ISC is competitive with fluorescence. Host rigidity is also important for suppressing deactivation of $\ket{S_1}$ by IC \cite{Nijegorodov1997}.

Pentacene itself has also been subject to chemical modification, including perdeuteration, which slows triplet decay by reducing vibrational coupling, and suppresses hyperfine-mediated spin dephasing resulting in narrowing of the spin transition line width \cite{Khler1998, ong1995deuteration}. A complementary strategy is heteroatom substitution, where replacing carbon atoms for nitrogen atoms enhances ISC by increasing vibrationally-mediated spin-orbit coupling through non-bonding orbital contributions, though at the cost of a significantly broadened inhomogenous line width due to $^{14}$N-hyperfine coupling \cite{bogatko2016molecular, attwood2023n}. Nevertheless, for 6,13-diazapentacene (DAP), which was previously synthesised as a more stable alternative to pentacene \cite{Kouno2019}, these changes significantly reduced the maser threshold \cite{ng2023move} and improved the spin-dependent photoluminescence contrast \cite{mann2025chemically}. However, the reduced $\ket{T_x}$ lifetime leads to an accumulation of the redundant $\ket{T_z}$ state, rendering DAP unsuitable for CW-masing. Overall then, whilst a significant body of work has investigated maser-relevant host-guest materials, these have been largely conducted under cryogenic conditions, or have not considered both the spin and optical properties, making defining the best direction for pursuing new quantum materials unclear.

In this work, we tackle this issue by using transient electron paramagnetic resonance (trEPR) and pulsed optically-detected magnetic resonance (ODMR) to systematically investigate the room-temperature spin dynamics of pentacene, here in referred to as Pc, doped into a series of host matrices: \textit{para}-terphenyl (Pc:PTP), perdeuterated \textit{para}-terphenyl (Pc-d$_{14}$:PTP-d$_{14}$), naphthalene (Pc:NAP), and picene (Pc:PIC). We also evaluate their performance in room-temperature maser devices and rationalise their performance in terms of both the host and guest structure and gauge their potential as media for developing CW masers. We demonstrate that the performance of Pc-host masers is strongly dependent on host-induced changes to Pc relaxation dynamics, which are themselves a product of changes in the nuclear spin-bath, molecular packing and vibrational coupling. Incorporation into maser devices reveals that Pc-d$_{14}$:PTP-d$_{14}$ leads to a more efficient maser, requiring only half of the optical pulse energy of its protiated analogue. In contrast, Pc:NAP and Pc:PIC struggle to mase, which can be rationalized by reduced spin polarization, triplet yield, and broader spin-transition line widths. These findings demonstrate host-guest engineering as a route to balance triplet generation, spin coherence and line width for optimised maser performance.

\section{Results and Discussion}

\begin{figure}[ht!]  
  \centering
  \includegraphics[width=\columnwidth]{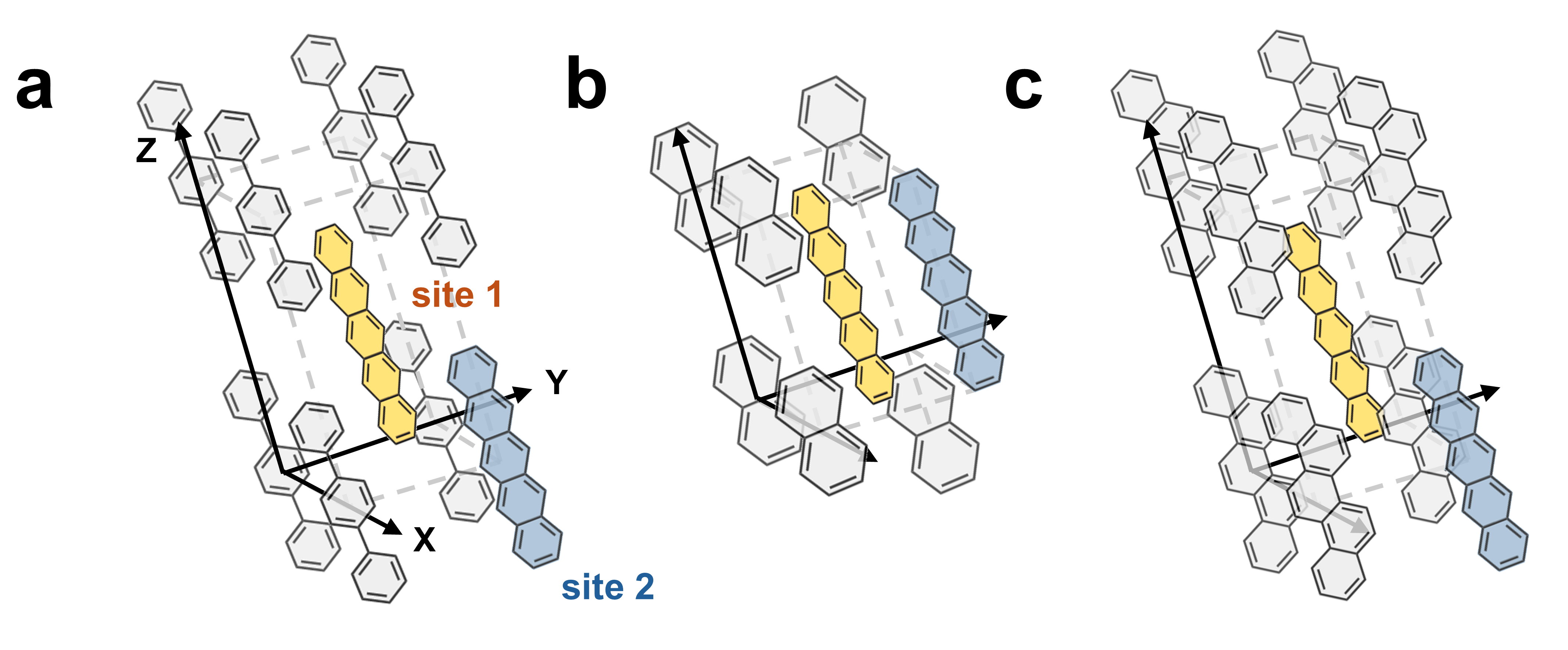}
  \caption{\textbf{Representative crystal structures of the host-guest systems.} Each system (a) Pc:PTP, (b) Pc:NAP and (c) Pc:PIC, adheres to a monoclinic crystal structure. Two possible crystallographic substitution sites for pentacene are shown in blue and yellow \cite{abrahams1949crystal,de1985structural}.}
  \label{packing}
  \vspace{2mm} 
  \hrule
  \vspace{1mm}
\end{figure}

Pc:PTP, Pc-d$_{14}$:PTP-d$_{14}$ and Pc:PIC were doped with nominal concentrations of 0.1 mol\% Pc, remaining below the regime where intermolecular interactions and aggregation become apparent. For trEPR experiments with Pc:NAP, we initially opted for a concentration of 0.01\% to avoid aggregation of Pc. To increase the spin concentration for maser and ODMR experiments, the pentacene loading was increased to 0.04 mol\% without any obvious pentacene aggregation in the crystal. These molar doping concentrations result in similar guest densities across the series with $\sim2.1 \times$ 10$^{24}$ m$^{-3}$ for 0.04 mol\% Pc:NAP, compared with $\sim3.3 \times 10^{24}$ m$^{-3}$ for 0.1 mol\% Pc:PTP and $\sim2.9 \times 10^{24}$ m$^{-3}$ for 0.1 mol\% Pc:PIC, placing the molecular spin reservoir well above the parts-per-million-level defect concentrations typically used in NV diamond masers \cite{jin2015proposal, breeze2018continuous, Sherman2022, Day2024}. For Pc:PIC, we routinely observed the separation of a small amount of black material following crystal growth which is likely due to some pentacene degradation resulting from the increased melting point of picene at 367$^o$C, compared with those of NAP (80$^o$C) and PTP (213$^o$C). Therefore, for subsequent experiments, we selected the best-looking crystal segments (see Figure S1). Within each host, Pc molecules are known to occupy two magnetically inequivalent crystal sites that arise due to the herringbone packing structure (Figure \ref{packing}).  

\subsection{Optical Characterisation}

\begin{figure*}[ht!]    
  \centering
  \includegraphics[width=0.65\textwidth]{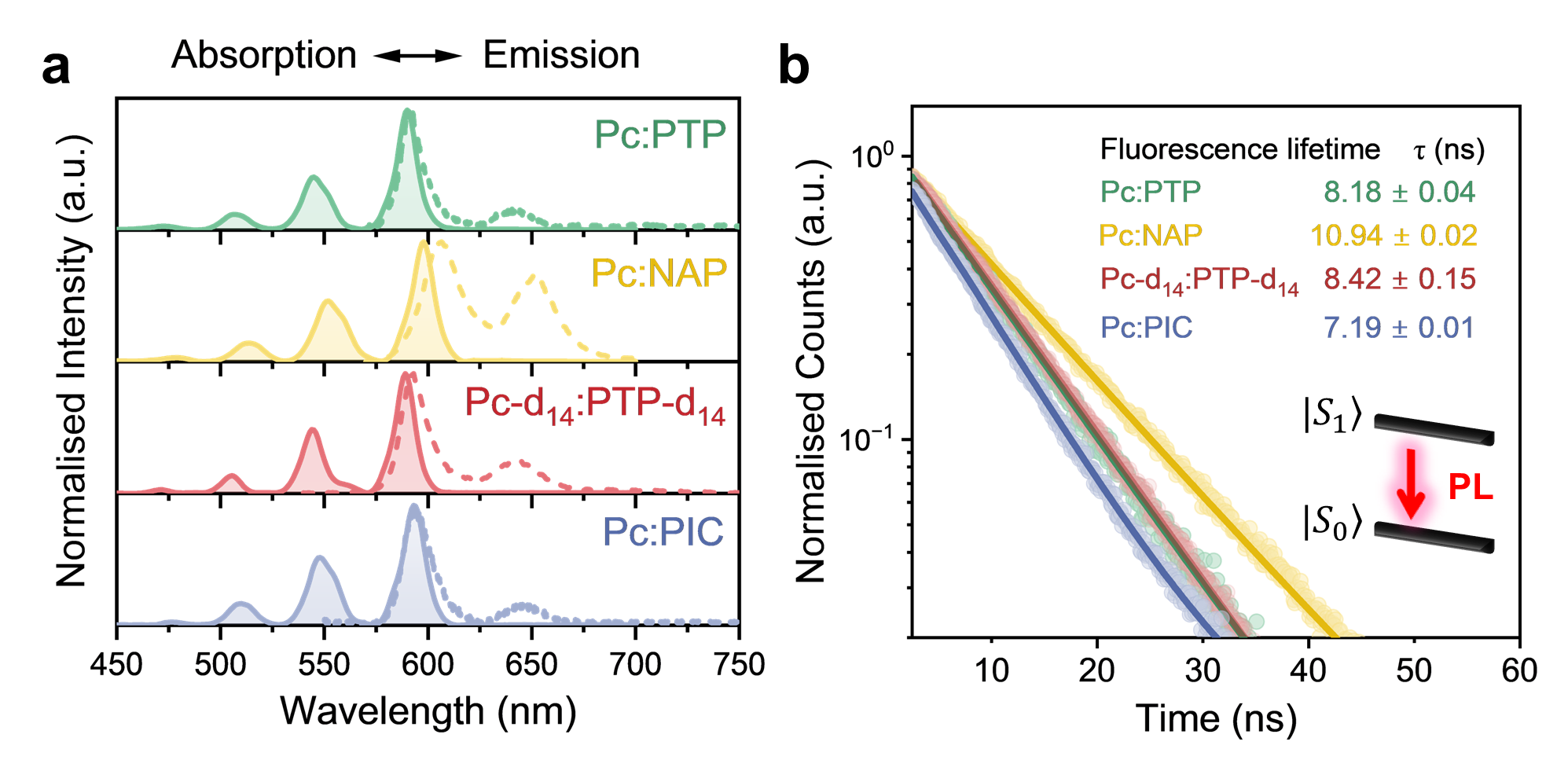}
  \caption{\textbf{Optical characterisation of Pc host–guest crystals measured at room temperature.} (a) Normalised UV–vis absorption (solid lines) and emission spectra ($\lambda_{ex}$ = 530 nm, dashed lines) for Pc:PTP (green), Pc:NAP (yellow), Pc-d$_{14}$:PTP-d$_{14}$ (red), and  Pc:PIC (blue). (b) Normalised time-correlated single photon counting (TCSPC) measurements of the fluorescence decay with inset weight-averaged fluorescence lifetimes.}
  \label{Optical}
  \vspace{2mm} 
  \hrule
  \vspace{1mm}
\end{figure*}

\begin{table*}[ht!]
	\centering
	\caption{X-band EPR fitting results of ZFS \textit{D} and $\lvert \textit{E} \rvert$ values, initial triplet population ratios from the highest sublevel $\ket{T_x}$ to the lowest sublevel $\ket{T_z}$ and spin decoherence time $T_{2,Rabi}$ of all host-guest systems in this study.}
	\footnotesize
	\begin{tabular}{lllllll}
	\hline
	Systems & \textit{D} / MHz & $\lvert \textit{E} \rvert$ / MHz & $P_x:P_y:P_z$ & $T_{2,Rabi}$ / $\mu$s @ LF & $T_{2,Rabi}$ / $\mu$s @ HF\\
	\hline
	Pc:PTP & 1394.8 $\pm$ 3.8 & 53.9 $\pm$ 1.59 & 0.76:0.16:0.08 & 3.40 $\pm$ 1.80 & 7.57 $\pm$ 0.83 \\
	Pc:NAP & 1383.4 $\pm$ 5.8 & 49.3 $\pm$ 2.71	& 0.69:0.20:0.11 & 0.77 $\pm$ 0.01 & 2.07 $\pm$ 0.02\\
	Pc-d$_1$$_4$:PTP-d$_1$$_4$ & 1389.8 $\pm$ 11.5 & 51.5 $\pm$ 5.10 & 0.82:0.18:0.00 & 6.96 $\pm$ 0.35 & 12.86 $\pm$ 1.79\\
	Pc:PIC & 1378.2 $\pm$ 3.6 & 53.2 $\pm$ 1.36 & 0.62:0.20:0.18 & 6.40 $\pm$ 3.43 & 12.77 $\pm$ 2.66\\
	\hline
	\label{EPRTABLE}
    \end{tabular}
\end{table*}

We began our study by clarifying the impact of different hosts on the optical properties of pentacene using steady-state absorption and fluorescence spectroscopy. Compared to Pc:PTP, the lowest energy singlet transition ($\ket{S_0} \rightarrow \ket{S_1}$) is modestly red-shifted for Pc:NAP ($\lambda_{max} \approx$ 598) and Pc:PIC ($\lambda_{max} \approx$ 592 nm), and blue-shifted for Pc-d$_{14}$:PTP-d$_{14}$ ($\lambda_{max} \approx$ 587 nm) (Figure \ref{Optical}a). This blue shift has previously been explained as arising from an increase in the frequency of C–D stretching modes \cite{kohler1998isotopomer}, whilst we assign the red shift seen with Pc:NAP and Pc:PIC to a stabilisation of the lowest unoccupied molecular orbital (LUMO) by enhanced $\pi-\pi$ interactions with the host, compared to PTP \cite{Grimme2008}. These $\pi-\pi$ interactions may also enhance the charge transfer interaction responsible for red-shifting the absorption of Pc:PTP compared to the gas phase pentacene \cite{Charlton2018, Bertoni2022}.

Since all of the hosts studied here are non-polar, the transition dipole moment that governs the rate of $\ket{S_1} \rightarrow \ket{S_0}$ relaxation is unlikely to change significantly. Therefore, differences in the $\ket{S_1}$ lifetime, which may be gauged by the fluorescence lifetime ($\tau_f$) are expected to reflect changes in the efficiency of ISC ($\frac{1}{\tau_f}=k_{\mathrm{r}}+k_{\mathrm{ISC}}+k_{\mathrm{IC}}$, where $k_{IC}$ represents the rate of IC). To probe these modified dynamics, we employed time-correlated single photon counting (TCSPC) to determine $\tau_f$ for each material (Figure \ref{Optical}b). All of these measurements were best fitted using a monoexponential decay (see ESI for details). For Pc:PTP, $\tau_f$ was found to be $\approx$ 8.18 ns, similar to 8.42 ns for Pc-d$_{14}$:PTP-d$_{14}$, and consistent with the reported values and established ISC yield of 62.5$£$\% \cite{patterson1984intersystem, yang2024bulk}. Pc:NAP exhibits a more gradual decay of $\sim11$ ns, indicating a reduced $\kappa_{ISC}$ which is again consistent with the material's previously estimated ISC yield of 40$£$\% \cite{takeda2002zero}. Interestingly, Pc:PIC exhibits the shortest $\tau_f$ (7.19 ns), which is nevertheless significantly longer than reported values for Pc:PIC thin-films \cite{Toccoli2018}. Compared to the reported values, we assign the difference to the absence of pentacene aggregates in Bridgmann-grown crystals which would otherwise accelerate $\ket{S_1}$ decay through multiexciton processes \cite{Lubert-Perquel2018, Zeiser2021}. 

Overall then, these data point to picene potentially facilitating the most efficient ISC. This is somewhat surprising, since in a rigid host such as picene, the amplitude of OOP-vibrations that promote ISC in pentacene should be impeded. This suggests instead that $\ket{S_1}$ and $\ket{T_2}$ of pentacene are even closer in energy within the picene host, compared to other systems \cite{corval1994resonant}. Changes to local vibrational modes that influence ISC dynamics are expected to be detectable either by a frequency shift of such modes in pentacene's vibrational spectrum, or in the relative spin populations of the triplet sublevels. Unfortunately, our attempts to conclusively identify pentacene vibrational modes using Raman spectroscopy were unsuccessful (see Figure S3). This is not surprising since pentacene comprises, at most, 0.1 mol\% of the bulk crystal. However, we note that similar OOP vibrations occur at significantly higher frequencies in NAP and PIC compared to PTP, suggesting that pentacene will adopt a more planar configuration in these hosts (see Table SII). Therefore, to probe the ISC dynamics further, we sought to determine the room temperature triplet sublevel populations using X-band trEPR spectroscopy. 

\subsection{X-band Transient EPR}

\begin{figure}[ht]
  \centering
  \includegraphics[width=\columnwidth]{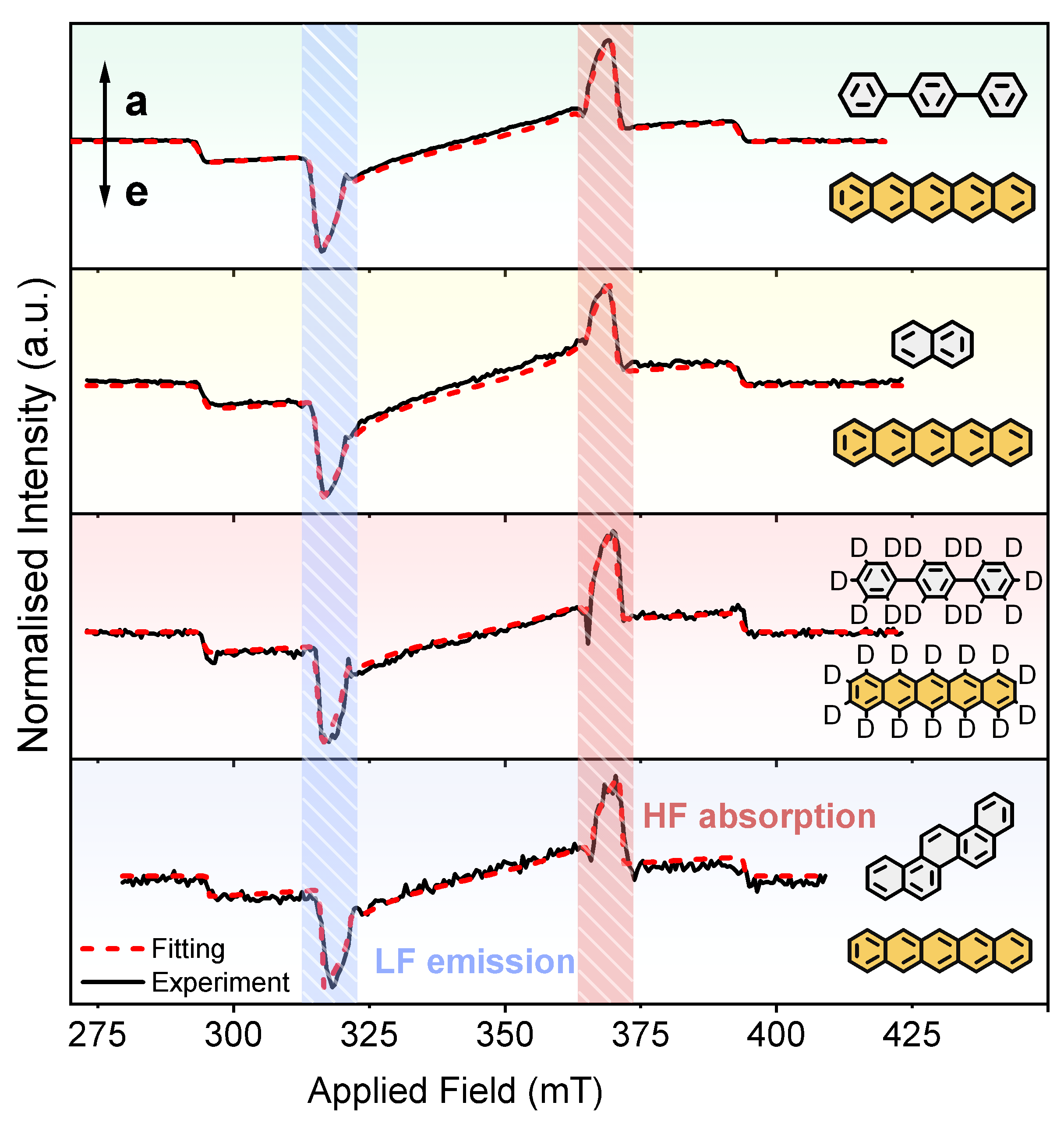}
  \caption{\textbf{Room-temperature trEPR experiments of all host-guest mixtures.} From top to bottom: experimental (black) and fitted (red) EPR spectra of 0.1\% Pc:PTP, 0.01\% Pc:NAP, 0.1\% Pc-d$_{14}$:PTP-d$_{14}$ and 0.1\% Pc:PIC taken at t = 600 ns, representing the point of maximum signal intensity. 
  Each spectrum was recorded using 590 nm light excitation, except for Pc:PIC which used 530 nm.}
  \label{All_EPR}
  \vspace{2mm} 
  \hrule
  \vspace{1mm}
\end{figure}

The powder spectra of all host–guest systems investigated in this study are shown in Figure \ref{All_EPR}. Their spectra exhibit a characteristic ISC triplet pattern consisting of an emissive–absorptive sequence (e, e, e, a, a, a). The ZFS parameters and initial relative triplet sublevel populations were obtained by fitting the data to a spin-Hamiltonian with both ZFS and Zeeman interactions (see ESI for details, summarised in Table \ref{EPRTABLE}). Across all host–guest systems, the ZFS parameters remain close to \textit{D} = 1.38 GHz and $\lvert \textit{E} \rvert$ = 50 MHz, with Pc:PIC exhibiting the largest deviation of 6 MHz (relative to Pc:PTP), pointing to weak perturbations of the local spin–spin interaction by the surrounding lattice. 

More substantial host dependence is observed in the instantaneous triplet sublevel populations. Pc:PTP exhibits a pronounced polarisation toward the $\ket{T_x}$ state ($P_x$:$P_y$:$P_z$=0.76:0.16:0.08), typical of ISC between $\pi\pi^*$-states. Upon deuteration, the Pc-d$_{14}$:PTP-d$_{14}$ system displays a detectably stronger population bias towards $\ket{T_x}$ and $\ket{T_y}$ ($P_x$:$P_y$:$P_z$=0.82:0.18:0.00). While earlier studies of Pc:PTP reported little to no change in the relative triplet sublevel populations upon deuteration \cite{ong1995deuteration}, ISC to $\ket{T_z}$ should become even less efficient due to an increase in the frequency of C-D bending vibrations (compared to C-H in Pc:PTP) likely responsible for coupling $\ket{S_1}$ to $\ket{T_z}$ \cite{corval1994resonant, Metz1972}. However, the best fit obtained for the Pc-d$_{14}$:PTP-d$_{14}$ triplet spectrum does not perfectly reproduce the experimental trace and so the populations are subject to some uncertainty (see ESI, Figure S5). In contrast, both Pc:PIC and Pc:NAP exhibit larger relative populations in the $\ket{T_y}$ and $\ket{T_z}$ states with $P_x$:$P_y$:$P_z$=0.62:0.20:0.18, and 0.69:0.20:0.11, respectively. These results demonstrate that, as expected, ISC to $\ket{T_x}$ and $\ket{T_y}$ in pentacene is less effective in NAP and PIC hosts, while relative coupling into $\ket{T_z}$ becomes more competitive.

Interestingly, we also observed Rabi oscillations in the time-domain traces for each material (Figure S6-S9). Measuring these oscillations under various microwave powers and fitting them to zero-order Bessel function, enabled the extraction of estimations for the spin decoherence time, $T_{2,Rabi}$ for the most prominent low-field (LF) and high-field (HF) features (see Table \ref{spin-dynamics-fit-results}, see ESI section EPR Rabi Oscillations for details). Pc:PIC and Pc-d$_{14}$:PTP-d$_{14}$ exhibited the longest $T_{2,Rabi}$, which can be attributed to reduced hydrogen density near the triplet $\pi-\pi^*$ orbitals compared to NAP and PTP. The difference between LF and HF values has previously been explained by the magnetic field dependence of hyperfine-mediated decoherence \cite{kouskov1995pulsed}.

\subsection{Quantifying Triplet Spin Dynamics with ODMR}

\begin{figure*}[hbt]
    \centering
    \includegraphics{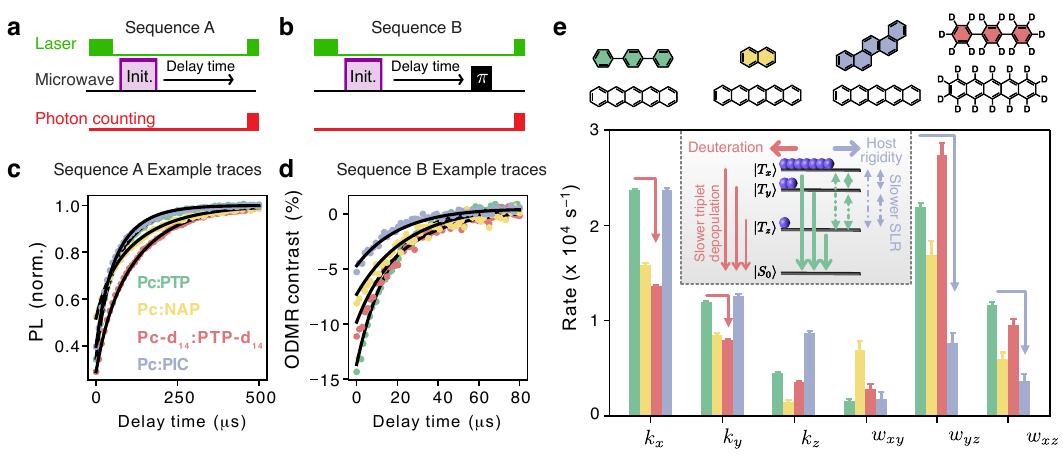}
    \caption{\textbf{Triplet spin dynamics from optically detected relaxation measurements.} (a, b) Pulse sequences A and B used to determine the dynamics (see Figure S13 for details of the pulse sequences, including the control and microwave initialisation sequences). (c, d) Example relaxation curves for Pc:PTP (reproduced from ref. \cite{mann2025chemically}), d\textsubscript{14}-Pc:d\textsubscript{14}-PTP, Pc:NAP and Pc:PIC recorded using \textbf{a} sequence A (Init 1.), and  \textbf{b} sequence B (Init 1. $|T_x \rangle \leftrightarrow |T_z \rangle$ readout), along with global fits (black lines). (e) Best-fit values of the spin dynamics rates. The arrows and inset illustrate the most significant differences between the materials.}
    \label{relaxation}
    \vspace{2mm} 
    \hrule
    \vspace{1mm}
\end{figure*}

\begin{table*}[ht]
\caption{Zero-field pulsed ODMR fitting results of triplet depopulation and spin-lattice relaxation rates of all host-quest systems in this study.}
\begin{tabular}{ccccccc}
\hline
&\multicolumn{6}{c}{Rate constant ($\times 10^4 \text{ s}^{-1}$)} \\  
&$k_x$ & $k_y$ & $k_z$ & $w_{xy}, w_{yx}$ & $w_{yz}, w_{zy}$ & $w_{xz}, w_{zx}$  \\ 
\hline
Pc:PTP \cite{mann2025chemically}           &$2.37 \pm 0.01$ & $1.20 \pm 0.01$ & $0.45 \pm 0.01$ & $0.15 \pm 0.03$ & $2.19 \pm 0.05$ & $1.17 \pm 0.03$ \\ 
Pc:NAP                                             &$1.59 \pm 0.02$ & $0.85 \pm 0.03$ & $0.15 \pm 0.02$ & $0.7   \pm 0.1$ & $1.7  \pm 0.2$  & $0.60 \pm 0.07$ \\
Pc-d\textsubscript{14}:PTP-d\textsubscript{14}     &$1.37 \pm 0.01$ & $0.80 \pm 0.01$ & $0.35 \pm 0.01$ & $0.28 \pm 0.05$ & $2.7  \pm 0.1$  & $0.96 \pm 0.06$ \\ 
Pc:PIC                                             &$2.37 \pm 0.02$ & $1.26 \pm 0.02$ & $0.87 \pm 0.02$ & $0.18 \pm 0.07$ & $0.8  \pm 0.1$  & $0.37 \pm 0.07$ \\ 
\hline
\end{tabular}
\label{spin-dynamics-fit-results}
\end{table*}

We next sought to determine the spin relaxation parameters relevant for quantum devices such as masers at zero-applied magnetic field (ZF). Previous investigations have used ZF-EPR spectroscopy to determine the triplet sublevel spin-lattice relaxation and decay rates \cite{ong1995deuteration, takeda2002zero, Wu2019UnravelingField, ng2023move, attwood2023n}. However, the narrow bandwidth of microwave components can make EPR a cumbersome technique to triage different materials. More recently, we have demonstrated that broadband, room-temperature pulsed ODMR can be a robust tool for determining the full suite of electron spin relaxation parameters for excited state triplets \cite{mann2025chemically}. This pulsed ODMR approach utilizes a large set of distinct time traces (22 in total; see pulse sequences in Figures \ref{relaxation}a, \ref{relaxation}b and S13) and the sensitivity of ODMR contrast to triplet kinetics to disentangle the spin-lattice relaxation and triplet decay processes that, in pentacene, occur on similar timescales.

After locating the CW-ODMR signals for all three triplet sublevel transitions (Figure S11), we quantified the triplet spin dynamics at ZF, extracting nine parameters: the triplet decay rates ($k_x$, $k_y$, $k_z$), the spin-lattice relaxation rates ($w_{xy}$, $w_{yz}$, $w_{xz}$), and the initial populations of the triplet sub-levels following laser excitation ($P_x$, $P_y$, $P_z$) \cite{mann2025chemically}. Figures \ref{relaxation}c and \ref{relaxation}d show example relaxation curves while all 22 curves and fits are shown in Figures S14-S16. The resulting parameters ($k_i$, $w_{ij}$, $P_i$) are shown in Figure \ref{relaxation}e and Tables \ref{spin-dynamics-fit-results} and SIV.

The spin-lattice relaxation rates retain the same ordering $w_{yz} > w_{xz} > w_{xy}$ across all host matrices. Remarkably, however, transitions involving the $\ket{T_z}$ state are significantly suppressed in Pc:PIC compared with Pc:PTP, (with $w_{yz}$ dropping from 2.19 to 0.8 $\times 10^4 \text{ s}^{-1}$ and $w_{xz}$ dropping from 1.17 to 0.37 $\times 10^4 \text{ s}^{-1}$). This effect is also seen to a lesser extent with Pc:NAP, which also exhibits a significantly increased $w_{xy}$. Altogether, this behaviour is consistent with the \textit{ab initio} Redfield calculations by Miyokawa and Kurashige \cite{miyokawa2026electron}, which showed that molecular and lattice vibrations drive spin--lattice relaxation by modulating the ZFS tensor. For Pc, the large ($w_{yz}$) and ($w_{xz}$) rates were associated with phenyl-ring rocking about the (X)-axis and librations about the (Y)-axis, respectively \cite{miyokawa2026electron,Ong1993TemperatureCrystals}. The amplitude of similar librations would be significantly suppressed in the more rigid PIC and NAP host which are characterised by their fused phenyl-rings.  Finally, the four-fold increase in $w_{xy}$ in Pc:NAP compared to the other hosts can likely be attributed to enhanced librations along Pc's Z-axis, since the corresponding distortions would be less impeded by the hinging of its two neighbouring NAP molecules. Finally, we observe a comparatively minor influence of deuteration on the spin-lattice relaxation rates which is consistent with earlier studies \cite{Yu1984TimeSpectroscopy, ong1995deuteration}. 

The host matrix was also found to influence the triplet depopulation rates, with $k_i$ varying across Pc:PTP, Pc:NAP, and Pc:PIC (Figure \ref{relaxation}c, Table \ref{spin-dynamics-fit-results}). In particular, the NAP host leads to a reduction of all the depopulation rates, whereas for PIC, $k_x$ and $k_y$ are comparable to PTP. This divergence between Pc:NAP and Pc:PIC mirrors our earlier observation that Pc:PIC exhibits faster $\ket{S_1}$ to $\ket{T_2}$ ISC than Pc:NAP. Since PIC is a more rigid host and the ISC rate is determined by electronic and vibrational coupling, this may also suggest that more efficient triplet decay in Pc:PIC results from $\ket{T_1}$ lying closer in energy to $\ket{S_0}$. However, we also note that in addition to exhibiting the largest shift in |D|, the CW-ODMR line width of Pc:PIC is significantly broadened compared to other materials (\textit{vide infra}, Figure S11). We attribute this difference to the impaired ability of PIC to accommodate the change in pentacene's geometry that occurs following it's transition to the triplet state. Hence, triplet decay in Pc:PIC is also likely enhanced by vibrational coupling. These results demonstrate that it is possible to engineer materials that exhibit advantages of both reduced spin-lattice relaxation and fast triplet decay. 

Deuteration reduces $\ket{T_1} \rightarrow \ket{S_0}$ triplet depopulation rates ($k_i$), which are approximately $1.5\times$ slower for Pc-d\textsubscript{14}:PTP-d\textsubscript{14} compared with Pc:PTP (Figure \ref{relaxation}e, Table \ref{spin-dynamics-fit-results}). This can be attributed to the suppression of non-radiative decay due to the lower vibrational frequency of C–D bonds relative to C–H bonds \cite{Yu1984TimeSpectroscopy, ong1995deuteration}.

Finally, taking care to minimise laser and microwave power broadening, we next measured the true CW-ODMR line width of the $\ket{T_x} \leftrightarrow \ket{T_z}$ masing transition at ZF, which is a key parameter that influences maser performance (see Figure S17, equation 2). As expected, Pc-d$_{14}$:PTP-d$_{14}$ exhibits the narrowest line width of 1.4 MHz followed by Pc:PTP (1.75 MHz), Pc:NAP (1.8 MHz) and finally, Pc:PIC (7 MHz). The significantly broadened line width of Pc:PIC reveals a trade-off between reducing the spin-lattice relaxation at the cost of increasing the line broadening. 

To further understand the source line width broadening, we measured the optically detected Hahn-echo lifetime for each material (Figure S19). Pc:PIC exhibits the longest $T_2$ of 1.51 $\mu$s, consistent with earlier X-band estimations, pointing to reduced dynamic sources of decoherence compared to the other materials. This confirms that static factors like strain are responsible for Pc:PIC's broadened line width. We also find that deuteration has little effect on $T_2$ compared to protiated-pentacene in either the PTP or NAP host. Since deuteration is expected significantly slow the evolution of the spin bath and increase $T_2$, it can be hypothesised that each sample is subject to a common dominant source of decoherence. A likely candidate is therefore that dynamic electron dipole interactions between neighbouring triplet spins \cite{Ryan2025}. This is consistent with longest $T_2$s being exhibited by Pc:NAP and Pc:PIC samples which harbour reduced pentacene-densities. Overall, then since dipole interactions also contribute to static broadening in solids with inhomogeneously distributed spins, the observed line widths are limited by both hyperfine and concentration-dependent electron-dipole interactions \cite{Glasbeek1984}. This is supported by the significantly reduced line width of 0.55 MHz measured for a sample of Pc-d$_{14}$: PTP-d $_{14}$ prepared at a concentration of 0.01 \ mol/mol (Figure S18).

Overall then, the enhanced spin polarisation of $\ket{T_x}$ state, combined with the reduced spin-lattice relaxation of the $\ket{T_x} \leftrightarrow \ket{T_z}$ transition and narrow line widths indicate that Pc-d$_{14}$:PTP-d$_{14}$ will be the better candidate for maser applications so far. 

\subsection{Microwave Amplification by Stimulated Emission of Radiation}

\begin{figure*}[ht]
	\centering
	\includegraphics[width=\textwidth]{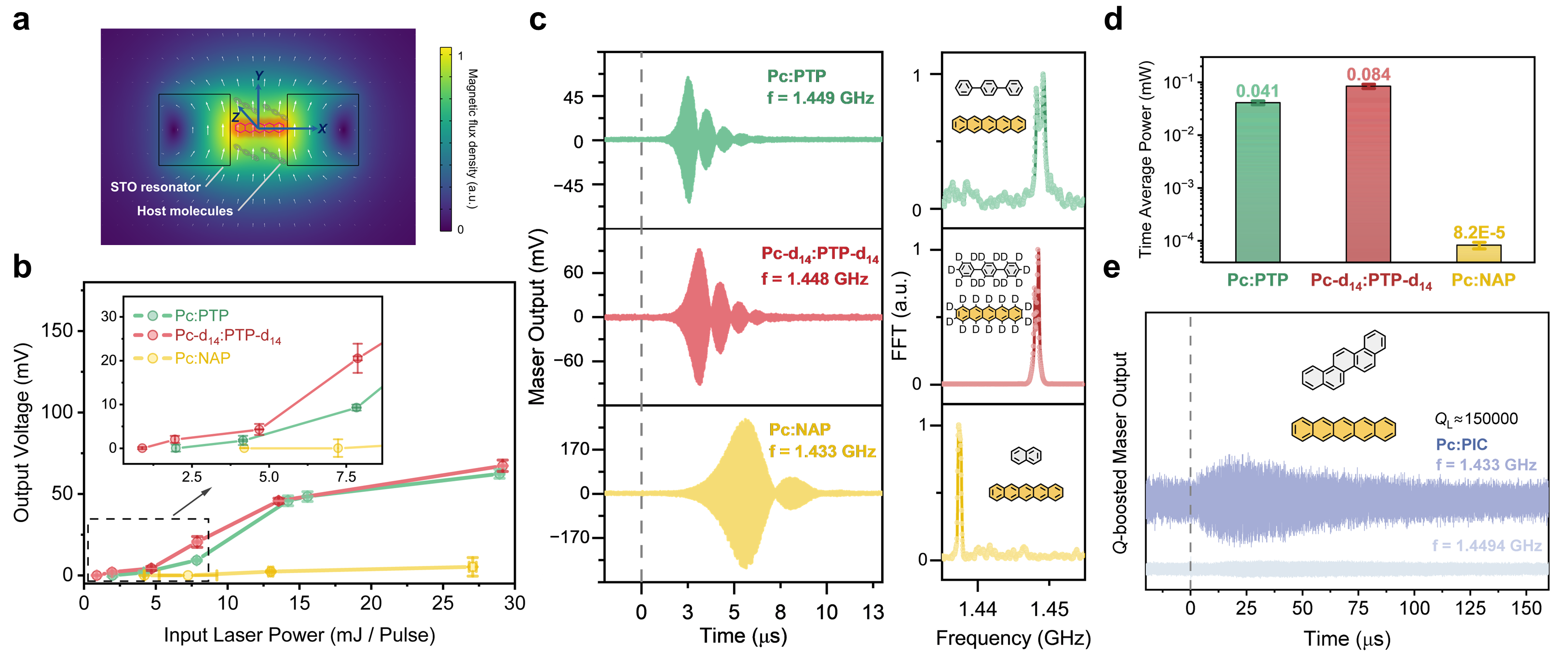}
	\caption{\textbf{Comparison of maser performance of different host-guest crystals.} (a) A two-dimensional simulation heat map of the magnetic energy density of the STO-loaded microwave cavity. The TE$_{01\delta}$ mode's a.c.~magnetic field drives $\ket{T_x}$ $\leftrightarrow$ $\ket{T_z}$ transitions. (b) Output voltage as a function of input laser power for three different host-guest systems: 0.1\% Pc:PTP (green), 0.1\% Pc-d$_1$$_4$:PTP-d$_1$$_4$ (red) and 0.04\% Pc:NAP (yellow). (c) Time-domain maser output signal recorded for three different host-guest systems at the maximum input laser energy. The grey dashed line indicates the laser trigger and the corresponding FFT of the steady-state maser spectra highlights host-dependent variations in line width. (d) Comparison of the real average peak maser powers in logarithmic scale for different systems from (c). (e) \textit{Q}-boosted maser signal for Pc:PIC at two different frequencies.}
	\label{Masing}
    \vspace{2mm} 
    \hrule
    \vspace{1mm}
\end{figure*}

To test these materials in a working device context, we constructed masers operating at room temperature using single crystals of each guest-host combination in a high-\textit{Q} strontium titanate (STO) resonator (Figure \ref{Masing}a). To build a useful maser, the spin dynamics of the candidate gain medium should facilitate the ability to mase with minimal optical pump energy, a measure called the maser threshold \cite{breeze2015enhanced}.

To determine the maser threshold, the pulsed maser signal was measured at increasing laser powers (Figure \ref{Masing}b). To optimise the sensitivity of our set up, care was taken to under-couple the resonator to maximise the quality-factor for each sample. This approach has previously been shown to significantly reduce the maser threshold, though at the expense of output extraction efficiency \cite{zollitsch2023maser}. Under these conditions, Pc-d$_1$$_4$:PTP-d$_1$$_4$ achieved the highest microwave output power and the lowest optical threshold at just 0.9 mJ, less than half that of Pc:PTP (2.4 mJ) and even less than DAP:PTP \cite{ng2023move, SophiaLong2026}, which is consistent with the reduced inhomogeneous line width and enhanced spin polarisation compared to other materials. Pc:NAP exhibits a significantly higher threshold of 4 mJ. The maser emission also started more slowly, and was significantly weaker, requiring another 40 dB microwave amplifier to enhance the signal (Figure \ref{Masing}d). Using above threshold laser energies, all maser signals exhibited Rabi oscillations that are characteristic of strong spin-photon coupling, with fast Fourier transform (FFT) of the time-domain oscillations revealing mode-splitting in the MHz regime (Figure \ref{Masing}c). Unfortunately, no maser signal was observed for Pc:PIC using the native quality-factor of the resonator. This apparent discrepancy can be rationalised by considering the effective line width of the $\ket{T_x} \leftrightarrow \ket{T_z}$ masing transition (7 MHz). To test whether this limitation could be overcome by enhanced microwave feedback, we constructed a \textit{Q}-boosted circuit to lower the maser threshold (Figure S22). The looped microwave circuit connects the input and output ports of the cavity via a low-noise amplifier, phase shifter and attenuator. This arrangement can form an external positive feedback loop that synthetically reduces the effective loss rate of the cavity by injecting an amplified version of the out-coupled microwave field back into the cavity. With an artificially increased \textit{Q} of $\sim150000$, a maser pulse was observed lasting about 50 $\mu$s (Figure \ref{Masing}e). To confirm this signal was due to the magnetic triplet state of Pc:PIC, a permanent magnet was introduced near the cavity to detune the microwave resonance, leading to the disappearance of the signal. 

\subsection{Toward Continuous Operation}

Finally, we turn to the prospect of building a CW-maser. The maser threshold is governed by the cooperativity, $\eta_{\mathrm{maser}}$, which quantifies the collective spin–photon interaction strength within the resonator and ultimately determines the noise figure of the maser as a microwave amplifier. It can be simply expressed as \cite{breeze2018continuous}:

\begin{equation}
\eta_{\mathrm{maser}} = 
\left( \frac{\mu_0 g^2 \mu_B^2}{h} \right) 
\eta_{\mathrm{mag.}} 
\frac{Q}{V_{\mathrm{mag.}}} 
\frac{\Delta N}{\Delta f} 
\label{equ_coop}
\end{equation}

where $h$ is the Planck's constant, $\eta_{\mathrm{mag.}}$ the magnetic filling factor, $\Delta N$ the population inversion between the $\ket{T_x}$ $\leftrightarrow$ $\ket{T_z}$ maser transition and $\Delta f$ the effective transition line width.

\begin{figure}[ht]  
  \centering
  \includegraphics[width=\columnwidth]{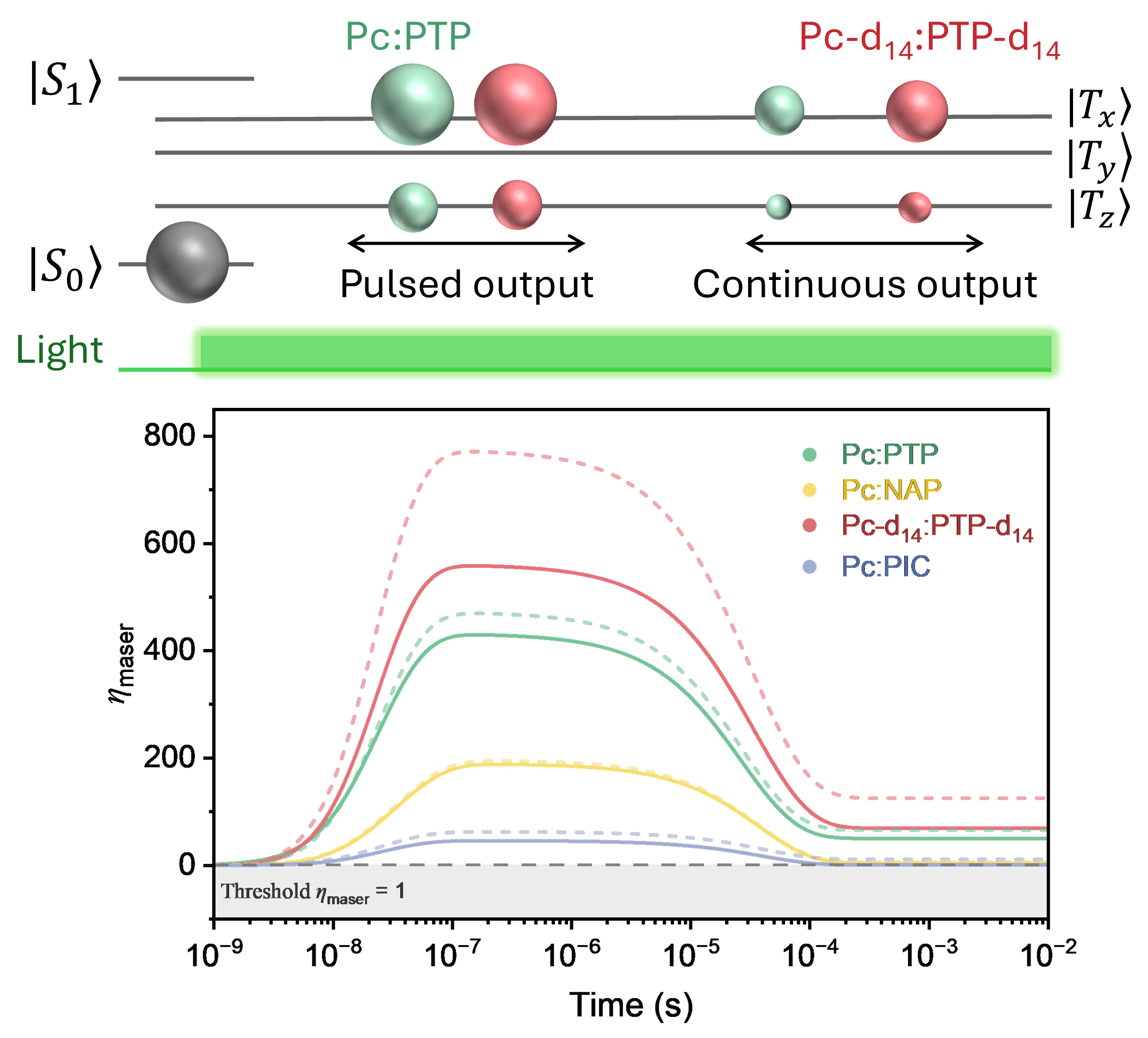}
  \caption{\textbf{Simplified energy level diagram and simulated time-dependent cooperativity $\eta_{\mathrm{maser}}$ for different host-guest candidates.} The $\eta_{\mathrm{maser}}$ is evaluated for the $\ket{T_x}$ $\leftrightarrow$ $\ket{T_z}$ maser transition using ODMR line widths and relaxation and depopulation rates. Solid and dashed curves correspond to ODMR- and EPR-derived triplet-sublevel populations respectively. The dashed horizontal line at the bottom marks the masing threshold, $\eta_{\mathrm{maser}}=1$.}
  \label{Coop_sim}
  \vspace{2mm} 
  \hrule
  \vspace{1mm}
\end{figure}

CW masing is achieved when $\eta_{\mathrm{maser}}$ exceeds unity, such that the collective gain overcomes the intrinsic cavity losses. As previously discussed, this is reliant on a narrow inhomogeneous line width, generating strong instantaneous spin polarisation, and enabling $\ket{T_y}$ and $\ket{T_z}$ to be recycled through the singlet state efficiently. To determine whether these changes have generated materials better suited to CW masing, we simulated the triplet kinetics and further simulated $\eta_{\mathrm{maser}}$ according to equation \ref{equ_coop} using our experimentally determined spin parameters (Figure \ref{Coop_sim}, see ESI for details). These simulations reveal that of these four materials, Pc-d$_{14}$:PTP-d$_{14}$ maintains the highest $\eta_{\mathrm{maser}}$ under continuous operation, which can be attributed to the larger spin polarisation and reduced emission line width of $\ket{T_x}\leftrightarrow\ket{T_z}$ transition.

\section*{Conclusions}

Quantum technologies that can be deployed under ambient conditions have tremendous potential to benefit society but are currently limited by their underlying active materials. For devices based on molecules, such as masers, these challenges may be resolved by chemically tuning their molecular spin properties. In this study, we have demonstrated that host–guest engineering provides a powerful route to systematically control maser properties in organic doped molecular crystals. By tailoring the lattice rigidity and spin–phonon coupling through host and isotopic substitution, we are able to reshape triplet sublevel populations, triplet depopulation pathways, and spin-lattice relaxation rates. Hosts such as para-terphenyl feature rotationally compliant scaffolds that can elastically accommodate the out-of-plane distortions of pentacene that promote spin-selective ISC to generate robust $\ket{T_x}\leftrightarrow\ket{T_z}$ population imbalance. However, the use of rigid hosts consisting of fused-rings, such as naphthalene and picene, reveals a complementary advantage of substantially suppressing spin-lattice relaxation and prolong triplet spin populations. This benefit is presently offset by reduced triplet polarisation and pronounced strain-induced line width broadening, which diminish maser performance. 
Our results therefore suggest an important materials-design trade-off for applications relying on efficient ISC: the ideal host should provide nuclear-spin dilution and sufficient rigidity to suppress detrimental relaxation pathways, while retaining enough lattice compliance to accommodate the guest molecule without excessive strain-induced line width broadening. The present results also suggest that no fundamental materials limitation precludes CW-maser operation in pentacene-based organic systems at room temperature and zero-applied magnetic field. However, to build a CW maser we expect that the heat load induced by laser excitation will still be a significant barrier due to thermal drift in the ZFS frequency \cite{lang2007dynamics, singh2025high} and frequency drift of the STO resonator \cite{wise2001structure}. Hence, further optimisation of resonator design \cite{xiang2025all}, cooling system, and excitation geometry \cite{wu2020invasive, liu2026scaleable} should be carefully considered. We anticipate that a CW-maser will become more accessible by using an invasive pumping scheme \cite{wu2020room, wu2020invasive}, and a lower concentration of pentacene due to the reduced line width and recent results demonstrating a lower pulsed maser threshold \cite{SophiaLong2026}. 

\section*{Experimental Section}
\textit{Sample Preparations}: Commercially purchased \textit{para}-terphenyl ($\geq$99\%, A14833, Alfa Aesar), naphthalene ($\geq$99\%, A13188, Alfa Aesar) and perdeuterated \textit{para}-terphenyl ($\geq$98\%, DLM-382-5, CIL) were zone-refined for at least 20 times before the crystal growth to achieve the optimum purity. Pentacene ($\geq$99.99\%, P2524, TCI) and picene ($\geq$99.9\%, P2207, TCI) were both sublimed under high vacuum before use, and pentacene-d$_{14}$ (98\%, CK Isotopes Ltd) was used as purchased.

\textit{Crystal Growth}: Single crystals used in this work were grown via the Bridgmann method \cite{Ai2017}. Different combinations of guests and hosts were mixed and thoroughly ground in a pestle and mortar, before being charged into a 6 mm borosilicate tube with one end sealed (100 x 6 mm, 1471153, Hilgenberg GmBH). This smaller tube was tube carefully put into a large borosilicate tube (8 mm OD) before being sealed under argon ($\sim900$ mbar). The bottom of the tube would then inserted into a furnace heated to 218$^o$C (for PTP), 80$^o$C (for NAP) or 375$^o$C (for PIC). The furnace was then moved along the length of the tube at 5 mm/hr. For Pc:PIC, we used a growth rate of 1 mm/hr due to the larger temperature gradient to reduce thermal stress. For ODMR measurements, the samples were grown directly inside 0.4 mm ID borosilicate rectangular capillary tubes (5004-100, CM Scientific) to ensure a flat geometry ideal for optical excitation. All crystals appeared uniformly pink under both light and UV illumination (see the Supporting Information for images). 

\textit{Materials Characterisations}: The UV–vis absorption spectrum was recorded for different crystals using an Agilent Cary 5000 UV–Vis–NIR spectrophotometer. The samples were excited using a 530 nm NanoLED diode laser (DD-532LN, 12489, HORIBA) that generated pulses shorter than 200 ps with an excitation density of 1 mWcm$^{-2}$ at a repetition frequency of 1 MHz during TCSPC measurements. 530 nm light sources were selected for excitation, and the photoluminescence signals were collected by a TBX-04 photomultiplier tube module detector. Raman spectroscopy was performed using a Renishaw (614Q37, inVia) Raman system with a confocal microscope containing a 785 nm, 300 mW diode laser (14483, 10785SR0100B, IPS). 

\textit{EPR Experiments}: Transient X-band EPR was performed at room temperature using Bruker X-band ELEXSYS E500T continuous-wave system equipped with Bruker ER4118X-MD-5W resonator and coupled with 20 Hz \textit{Q}-switched Nd:YAG laser (Spectra-Physics INDI-HG-20S) pumped broadband OPO (GWU, primoScan/ULD 120). The laser pulse energy was measured to be $\sim5$ mJ (pulse length 5-7 ns). Crystals were ground into fine powders and transferred to 4 mm O.D. Wilmad quartz (CFQ) EPR tubes. The powder spectrum fittings were performed with \textit{EasySpin} \cite{stoll2006easyspin} 6.0.10 toolbox.

\textit{ODMR Experiments}: ODMR measurements were performed at room temperature and zero field. Samples were excited using a \(520\)\,nm laser diode (either a Thorlabs LP520-SF15 for the CW-ODMR line width measurements or a Thorlabs LP520-SF40 for all other measurements), which was filtered using a \(520-40\)\,nm band-pass filter (Thorlabs FBH520-40). A dichroic mirror (Thorlabs DMLP550R) was used to direct the excitation laser to the sample through either an achromatic lens (Thorlabs AC254-030-AB) for CW-ODMR line width measurements of Pc:PTP, Pc:PIC, Pc-d\(_{14}\):PTP-d\(_{14}\) (Figures S17 and S18), an aspheric lens (C240TMD-B) for the CW-ODMR line width measurement of Pc:NAP (Figure S17), or a microscope objective (Zeiss 100$\times$/0.9 DIC M27) for all other measurements. Photoluminescence (PL) was collimated by the same lens, separated from the excitation through the dichroic, and passed through a 550\,nm long-pass filter (Thorlabs FELH0550). For CW-ODMR and Rabi measurements (Figures S11 and S12) the PL was coupled into a multi-mode fiber using an achromatic lens (Thorlabs AC254-030-AB) and sent to a photodetector (FEMTO, OE-200-SI). For the CW-ODMR line width measurements (Figures S17 and S18), the PL was collected in free space and focused onto a photodetector (FEMTO OE-200-SI) using an achromatic lens (Thorlabs AC254-030-AB). The PL signal from the FEMTO detector was fed into a lock-in amplifier (Stanford Research Systems, SR830) which was referenced to the microwave modulation frequency. For the triplet spin dynamics measurements, the PL was coupled into a single-mode fiber using an aspheric lens (Thorlabs A260TM-B) and sent to a single photon counting avalanche photodiode (APD, Excelitas SPCM-AQRH-14-FC). Voltage pulses from the APD were counted using a time-to-digital converter (Swabian Instruments TimeTagger 20). The laser diode was operated in CW-mode for CW-ODMR spectra, or pulsed through a pulse generator (Agilent 8114A) for all pulsed measurements. The microwave fields were generated from one of two signal generators (Stanford Research Systems SG396; and Windfreak SynthNV PRO), each of which was gated using a high isolation microwave switch (Minicircuits ZASWA-2-50DRA+). For the CW-ODMR line width measurements the microwave fields were generated using a signal generator (Rohde and Schwarz SMA100B for the Pc:PTP and \(0.1\%\) Pc-d$_{14}$:PTP-d$_{14}$, and Agilent E8257C for the \(0.01\%\) Pc-d$_{14}$:PTP-d$_{14}$, Pc:PIC, Pc:NAP) which was gated using a high isolation microwave switch (Minicircuits ZASWA-2-50DRA+).The microwave pulses from the two channels were combined using a power splitter (Minicircuits ZFSC-2-372-S+), amplified (Minicircuits ZHL-25W-272+) and then sent to either a microwave planar loop antenna with an internal diameter of 1 mm or printed circuit board co-planar waveguide. Microwave and laser pulses, and photon counting were synchronized by an arbitrary waveform generator (Swabian Instruments Pulse Streamer 8/2).

Relaxation measurements conducted with Sequence A and Sequence B used microwaves on resonant with the three triplet sublevel transitions, $|T_x\rangle \leftrightarrow |T_y\rangle$, $|T_y\rangle \leftrightarrow |T_z\rangle$ and $|T_x\rangle \leftrightarrow |T_z\rangle$ at 107.5\,Mz, 1.341\,GHz and 1.448\,GHz respectively for Pc-d$_{14}$:PTP-d$_{14}$, 105.3\,Mz, 1.333\,GHz and 1.439\,GHz respectively for Pc:PIC, and 107.4\,MHz, 1.342\,GHz and 1.449\,GHz for Pc:NAP. $\pi$ times for all transitions were determined via Rabi measurements (Figure S12). Measurements used a 5 $\mu$s optical initialisation pulse, a 0.5\,$\mu$s readout-out pulse, and in Sequence B, a 30\,$\mu$s readout delay. The repetition time was set to 600 $\mu$s to ensure complete relaxation to the singlet ground state between measurement repeats.

Each set of 22 relaxation measurements were simultaneously fit to the rate equation model as outlined in ref. \cite{mann2025chemically}.

\textit{Maser Setup}: \textit{Q} and coupling strength were determined to be $\sim3000$ and $\sim-0.1$ dB respectively for all materials, as quantified using an HP 8753A vector network analyser (VNA). The simulation of magnetic flux and field distribution of the cavity was performed using COMSOL Multiphysics$^\circledR$ \cite{multiphysics2020comsol}. The final voltage signal was captured at an input impedance of 50 $\Omega$ by a Keysight DSOX 6002A 6 GHz oscilloscope with a sampling rate of 5 GSas$^{-1}$.

\section*{Supporting Information}
Supplementary information for crystal photos, TCSPC fittings, Raman spectra, the spin-Hamiltonian for the pentacene triplet electron, trEPR Rabi oscillation fittings and explanations, ODMR relaxation traces, ODMR line widths, spin dynamics simulation details and \textit{Q}-boost maser experiment set-up can be found at [URL to be added].

\section*{Author contributions}
\textbf{Z.H.:} methodology, curation of UV/Vis, Raman, EPR, DFT, and Maser data, formal analysis, visualization, writing – original draft. 
\textbf{A.C.-S.:} CW-ODMR data curation, formal analysis, visualization. 
\textbf{I.N.:} supervision of EPR, writing – review and editing.
\textbf{Y.Y.:} investigation of \textit{Q}-boost Maser experiment, visualization. 
\textbf{M.O.:} conceptualization, supervision, writing – review and editing. 
\textbf{S.B.:} supervision, writing – review and editing, funding acquisition.
\textbf{S.K.M.:} curation of CW-ODMR and triplet kinetics data, formal analysis, visualization, writing – review and editing, funding acquisition. 
\textbf{M.A.:} conceptualization, investigation of crystal growth, supervision of the overall research, writing – review and editing, funding acquisition.

\section*{Conflicts of interest}
No conflict of interest to declare.

\section*{Data availability}
The data underlying this work are available at [URL to be added].

\section*{Acknowledgements}
The authors would like to acknowledge the EPSRC equipment funding for SPIN-Lab (EP/P030548/1). A.C-S and S.L.B. acknowledge support from UK Research and Innovation [grant number MR/W006928/1]. M.A also acknowledges support from the UK Research and Innovation [grant number EP/W027542/1]. S.K.M acknowledges support from a Glasgow Engineering Futures Fellowship. For the purpose of open access, the authors have applied a Creative Commons Attribution (CC BY) licence to any Author Accepted Manuscript version arising from this submission.


\bibliography{References}
\bibliographystyle{rsc}
\end{document}


\setcounter{section}{0}

\title{Supporting Information for: 
Host-guest Crystal Engineering Tailors the Room Temperature Spin Dynamics in Molecular Quantum Devices}

\author{Ziqiu Huang}
\affiliation{Department of Materials and London Centre for Nanotechnology, Imperial College London, Prince Consort Road, London, SW7 2AZ, UK}

\author{Angus Cowley-Semple}
\affiliation{James Watt School of Engineering, University of Glasgow, Glasgow, G12 8QQ, UK.}

\author{Irena Nevjestic}
\affiliation{Department of Materials and London Centre for Nanotechnology, Imperial College London, Prince Consort Road, London, SW7 2AZ, UK}

\author{Yifan Yu}
\affiliation{Department of Materials and London Centre for Nanotechnology, Imperial College London, Prince Consort Road, London, SW7 2AZ, UK}

\author{Mark Oxborrow}
\email{m.oxborrow@imperial.ac.uk}
\affiliation{Department of Materials and London Centre for Nanotechnology, Imperial College London, Prince Consort Road, London, SW7 2AZ, UK}

\author{Sam L. Bayliss}
\email{sam.bayliss@glasgow.ac.uk}
\affiliation{James Watt School of Engineering, University of Glasgow, Glasgow, G12 8QQ, UK.}

\author{Sarah K. Mann}
\email{sarah.mann@glasgow.ac.uk}
\affiliation{James Watt School of Engineering, University of Glasgow, Glasgow, G12 8QQ, UK.}

\author{Max Attwood}
\email{m.attwood@imperial.ac.uk}
\affiliation{Department of Materials and London Centre for Nanotechnology, Imperial College London, Prince Consort Road, London, SW7 2AZ, UK}

\maketitle
\tableofcontents
\clearpage

\section{Pictures}
\begin{figure}[H]
    \centering
    \includegraphics[width=0.7\linewidth]{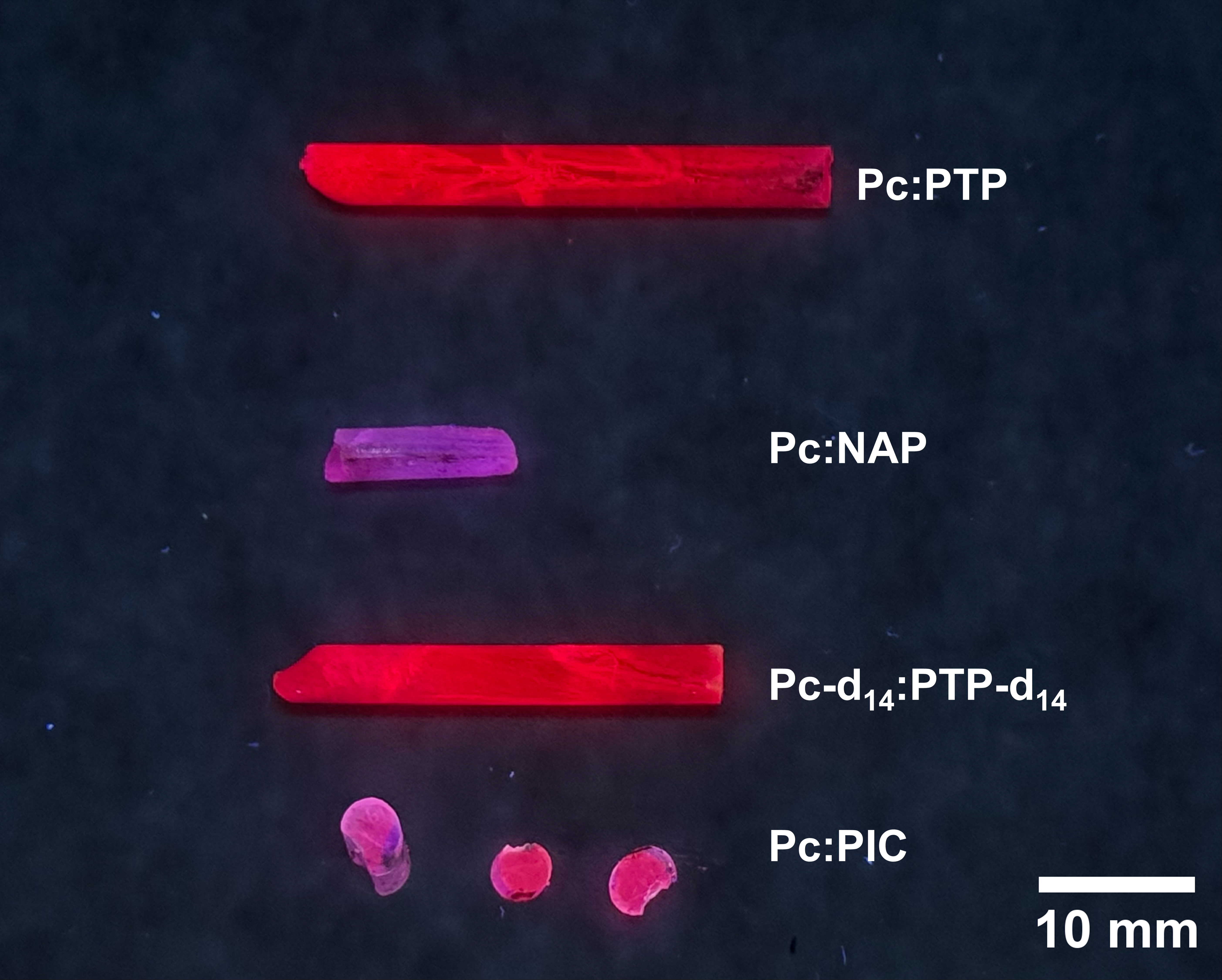}
     \caption{\textbf{Photographs of pentacene-doped host crystals under long-wave UV excitation: 0.1\% Pc:PTP, 0.04\% Pc:NAP, 0.1\% Pc-d$_{14}$:PTP-d$_{14}$ and 0.1\% Pc:PIC.} Pc:PIC sample appears fragmented into several smaller pieces due to the relatively high melting point of picene, which results in fracturing when the crystal is cooled.}
    \label{Crystal_picture}
\end{figure}
\clearpage

\clearpage
\section{Transient Fluorescence Spectroscopy}

The fluorescence lifetimes were determined by fitting the TCSPC decay 
curves using iterative reconvolution. The experimentally measured 
instrument response function (IRF) was convolved with a monoexponential decay 
according to:
\begin{equation}
I_{\mathrm{fit}}(t)=I_0\left[\mathrm{IRF}(t)\otimes\exp\left(-\frac{t}{\tau}\right)\right]+C\label{equ_tcspc}
\end{equation}
where $C$ is the constant background offset, $I_{0}$ is the 
pre-exponential amplitude, and $\tau$ is the fluorescence lifetime.

\begin{figure}[ht!]
    \centering
    \includegraphics[width=0.9\linewidth]{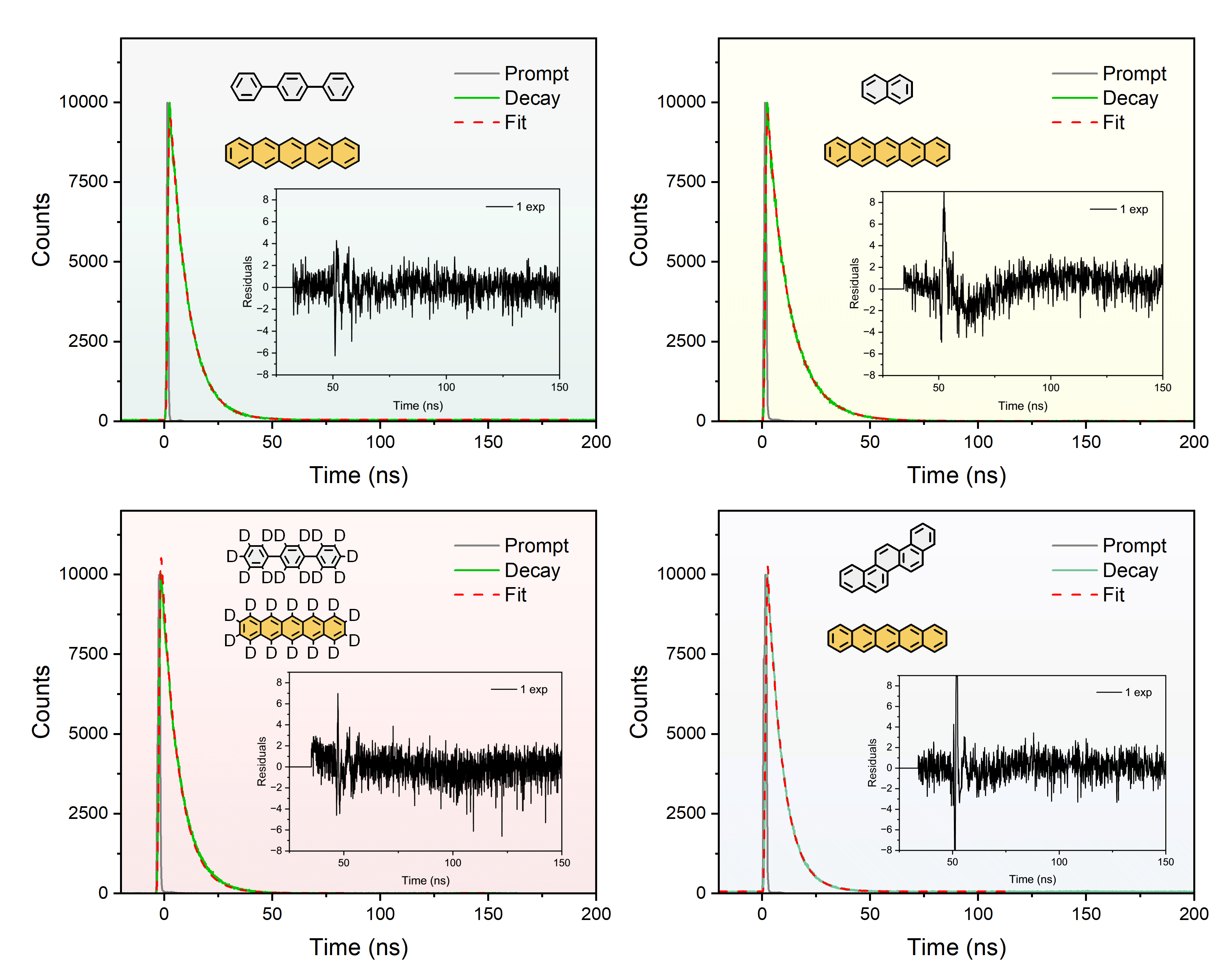}
    \caption{\textbf{TCSPC spectroscopy of all host-guest systems.} Insets are the residual traces after fitting.}
\end{figure}


\begin{table}[htbp!]
\centering
\caption{Summary of monoexponential fitting parameters for the TCSPC fluorescence decay measurement, which were fitted using a monoexponential model (Equation \ref{equ_tcspc}) convolved with the measured instrument response function.}
\label{tcspc_parameters}
\begin{tabular}{lcccc}
\hline
Sample &
$C$ &
$I_{0}$ &
$\tau$ (ns) &
$R^2$ \\
\hline
Pc:PTP &
$54.48 \pm 0.27$ &
$11455 \pm 21$ &
$8.18 \pm 0.04$ &
0.999 \\

Pc:NAP &
$7.29 \pm 0.08$ &
$10805 \pm 20$ &
$10.94 \pm 0.02$ &
0.998 \\

Pc-d$_{14}$:PTP-d$_{14}$ &
$2.08 \pm 0.13$ &
$11944 \pm 16$ &
$8.42 \pm 0.15$ &
0.999 \\

Pc:PIC &
$58.38 \pm 0.22$ &
$11808 \pm 26$ &
$7.19 \pm 0.01$ &
0.998 \\
\hline
\end{tabular}
\end{table}
\clearpage

\clearpage
\section{Raman Spectroscopy}

To quantify the relative spatial confinement imposed by each host lattice, we estimated an effective intermolecular length scale by converting the molecular volume per host molecule into an equivalent average intermolecular spacing: $d_{host}$ $\approx$ $(M/\rho N_A)^{1/3}$. This yields $d_{host}$ $\approx$ 5.9 \r{A} for NAP, 6.8 \r{A} for PTP and 6.9 \r{A} for PIC. This length scale provides a simple metric for comparing the relative packing density of the host matrices and therefore the degree of constraint imposed on the doped pentacene molecules. The smaller value for NAP suggests a more spatially confined environment, which may restrict large-amplitude out-of-plane (OOP) distortions of the embedded pentacene molecule. 

\begin{figure}[H]
 \centering
    \includegraphics[width=0.9\linewidth]{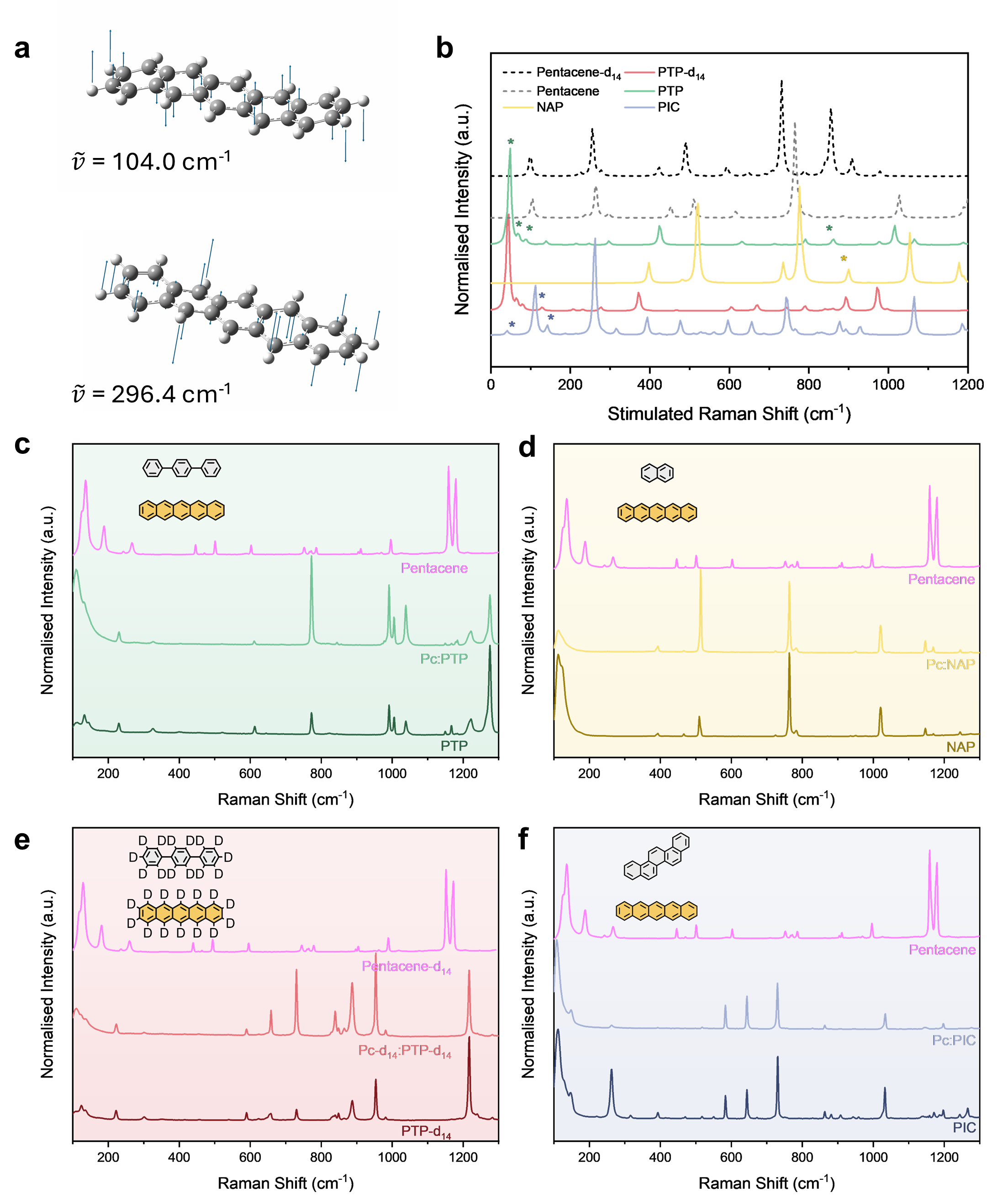}
    \caption{\textbf{Simulated and experimentally measured Raman spectroscopy of pentacene, pure hosts, and measured spectra of host-guest combinations.} (a) Atomic displacements of pentacene for two out-of-plane modes with enlarged displacement vectors. (b) DFT-computed Raman modes of isolated host and guest molecules in gas phase (dashed lines), whereas the marks show the peaks with a triplet-favouring vibration. (c-f) Experimental Raman spectra of all pure host and guest powder before and after doping. All spectra were taken under continuous illumination with a 785 nm laser.}
    \label{Raman}
\end{figure}

Density functional theory (DFT) calculations were performed using the \textit{Gaussian} package \cite{g16}. Molecular structures were initially built and pre-optimised in GaussView, before full geometry optimisation using the B3LYP exchange--correlation functional and the 6-31G(d,p) basis set. Geometry optimisations were followed by vibrational frequency calculations to confirm that the optimised structures corresponded to true minima. Two lowest OOP modes of pentacene are shown in Figure \ref{Raman}a. The calculated Raman spectra for each material are shown in Figure \ref{Raman}b, with OOP vibrations indicated by an asterisk. PTP displays OOP features that remain close to the pentacene OOP vibrations referenced by Kryschi \textit{et al.} \cite{kryschi1992vibronically}. Comparison between PTP and PTP-d$_{14}$ reveals the expected reduction in vibrational frequency for modes involving C–D (compared to C-H). In the PIC host, the Raman-active OOP features are less pronounced, indicating that these distortions are not strongly expressed in the measured Raman spectrum. Pc:NAP shows the largest deviation of the low-frequency OOP modes from their gas-phase positions. This trend is consistent with the smaller effective intermolecular spacing of NAP, which may lead to stronger steric confinement of the pentacene framework and consequently stiffen the OOP vibrational modes, shifting them to higher energies.

In addition to those low-frequency modes, pronounced differences are observed in the 850–950 cm$^{-1}$ region, where pentacene framework vibrations have recently been discussed as important modes for triplet-DNP-relevant spin polarisation \cite{sakamoto2023polarizing}. Again, Pc:NAP exhibits strong Raman bands that are shifted from the gas-phase pentacene reference positions, while Pc:PIC shows less clearly resolved features. This indicates that both host packing and isotope substitution perturb the normal-mode composition of the pentacene skeleton rather than merely shifting isolated molecular vibrations. Combined with the TCSPC results, these observations suggest that the reduced triplet yield observed in Pc:NAP may arise from the host-induced modification of the ISC-promoting vibrational modes. PIC does exhibit several low-frequency OOP modes, suggesting that it can support geometrically similar distortions. 

\begin{table}[htbp]
\centering
\caption{Summary of calculated Raman-active vibrational modes of the host molecules in the low-frequency and triplet-DNP-relevant spectral regions. The mode assignments are based on DFT-calculated vibrational modes.}
\label{tab:host_raman}
\renewcommand{\arraystretch}{1.2}
\resizebox{\textwidth}{!}{%
\begin{tabular}{llll}
\hline
Host & Peak / cm$^{-1}$ & Mode type & Main displacement \\
\hline

PTP & 
37.69, 88.2, 861.61 & 
$x$-axis distortion & 
Butterfly-like / OOP bending \\

PTP & 
69.41 & 
$y$-axis distortion & 
Twisting / torsional motion \\

\hline

NAP & 
397.11, 481.24, 899.37 & 
$x$-axis distortion & 
Butterfly-like / OOP bending mode \\

NAP & 
517.85 & 
$y$-axis distortion & 
Twisting / in-plane distortion mode \\

\hline

PIC & 
40.66, 111.37 & 
$x$-axis distortion & 
Butterfly-like / OOP bending \\

PIC & 
141.88 & 
$y$-axis distortion & 
Twisting / transverse distortion \\

PIC & 
876.21 & 
Aromatic ring deformation & 
Peripheral C--H motion / limited backbone distortion \\

\hline
\end{tabular}%
}
\end{table}

\begin{figure}
    \centering
    \includegraphics[width=\linewidth]{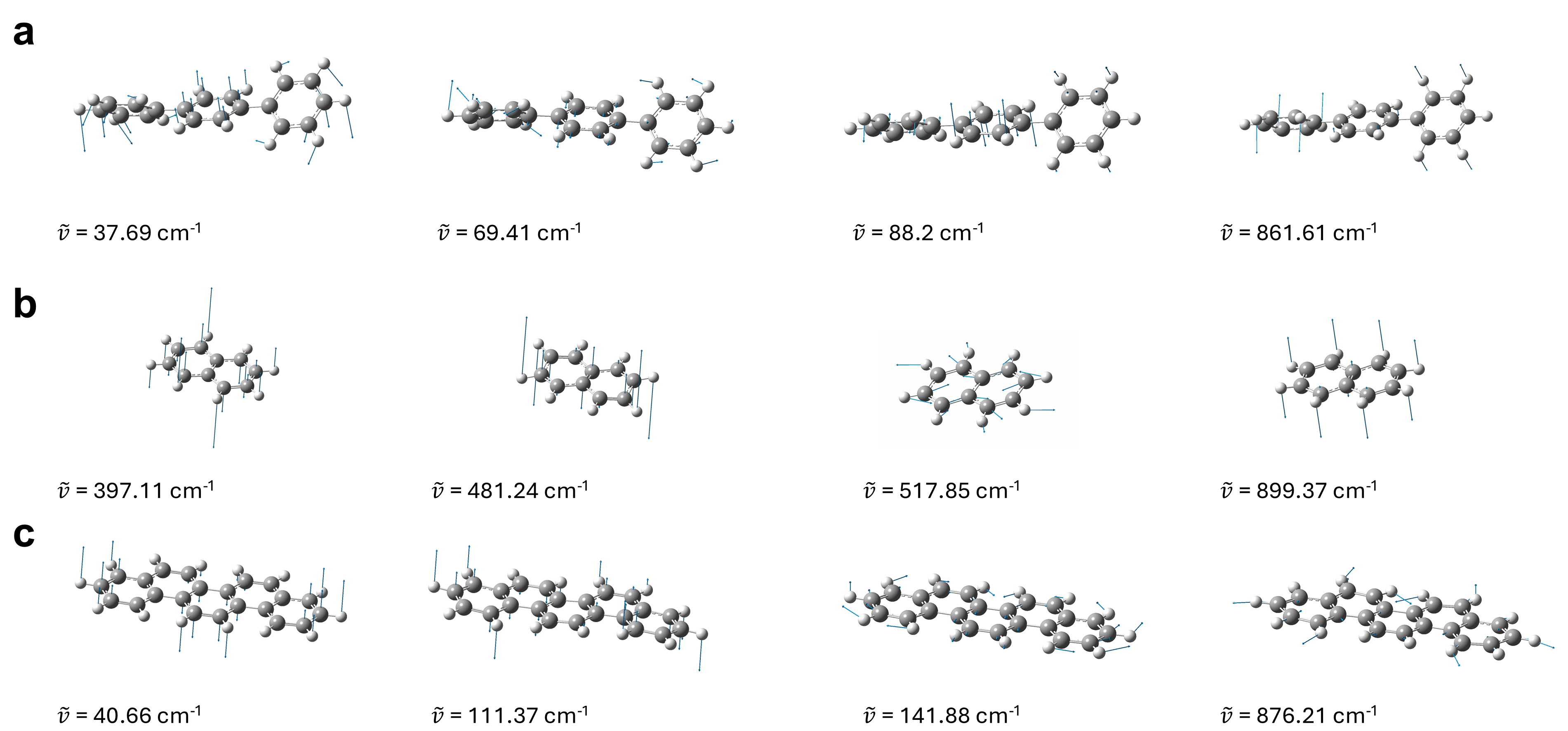}
    \caption{Selected DFT-calculated Raman-active vibrational modes of the host molecules (a) PTP, (b) NAP and (c) PIC.}
    \label{host_raman}
\end{figure}
\clearpage

\clearpage
\section{Additional EPR Spectroscopy}
The spin Hamiltonian used to describe the photoexcited triplet state of pentacene is given by:
\begin{equation}
H = D\left(S_z^2 - \frac{1}{3}S^2\right) + E\left(S_x^2 - S_y^2\right) + g_e \mu_B \mathbf{B}_0 \cdot \mathbf{S}.
\end{equation}
where $D$ and $E$ are the zero-field splitting (ZFS) parameters describing the anisotropic dipolar interaction between the two unpaired electrons in the triplet state. $g_e \approx 2.0023$ is the electron $g$-factor, $\mu_B$ is the Bohr magneton, $\mathbf{B}_0$ is the applied magnetic field, and $\mathbf{S}$ represents the total electron spin operator ($S=1$ for a triplet state). The first two terms describe the intrinsic zero-field splitting of the triplet manifold arising from electron--electron dipolar coupling, while the final term accounts for the Zeeman interaction in the presence of an external magnetic field.

\begin{figure}[htbp]
    \centering
    \includegraphics[width=\linewidth]{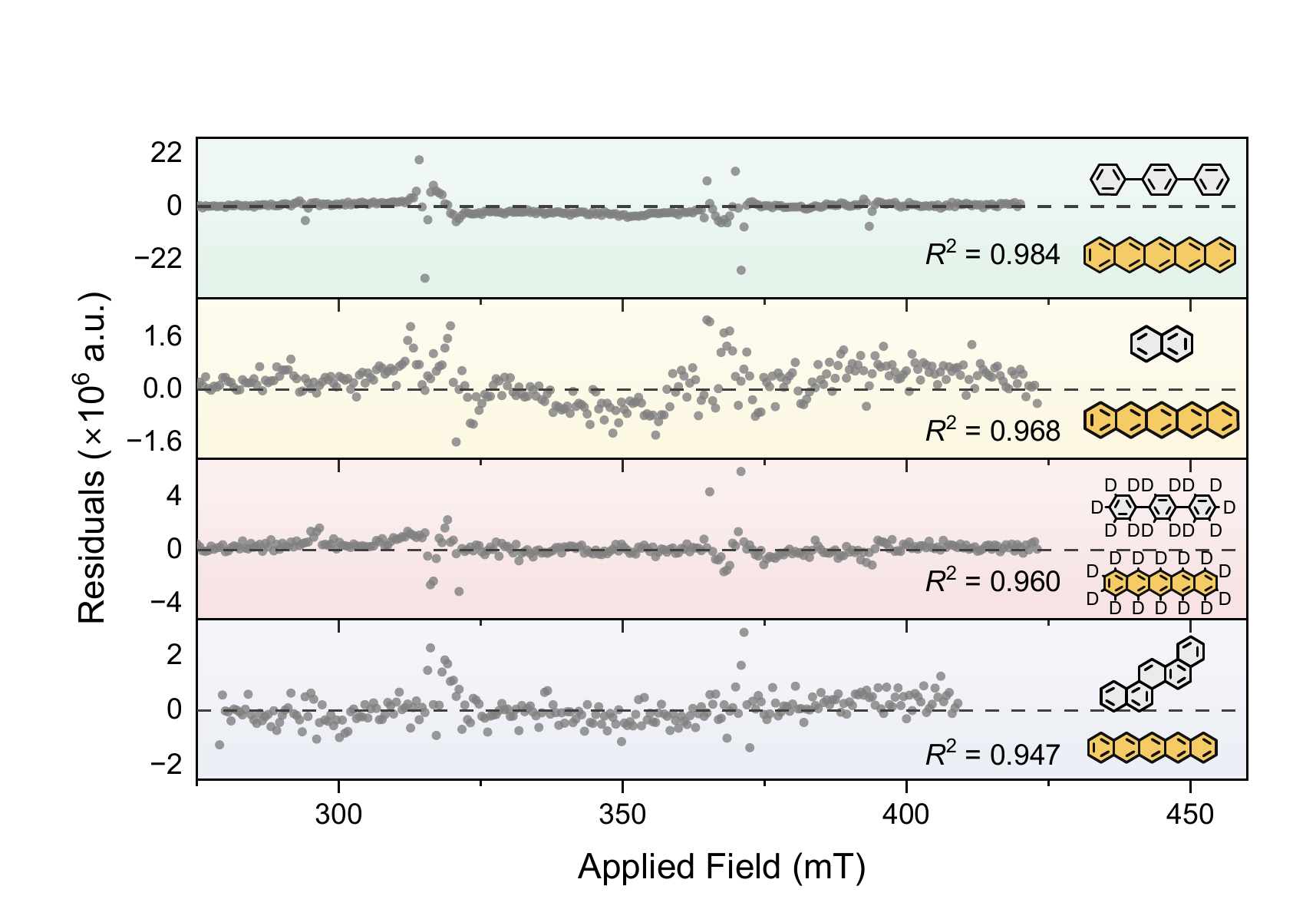}
    \caption{\textbf{Residuals of the simulated trEPR powder spectra for the host-guest systems.} Residuals were calculated as the difference between the experimental and simulated spectra over the fitted magnetic-field range.}
    \label{eprfttiingerror}
\end{figure}

To extract the spin Hamiltonian parameters, the experimental powder EPR spectra were simulated and fitted using the \textit{esfit} routine implemented in the \textit{EasySpin} \cite{stoll2006easyspin} package for MATLABR2024b. This approach allows the zero-field splitting parameters and the relative triplet sublevel populations to be determined by optimising the agreement between the simulated powder spectrum and the experimental data. The residual plots are shown in Figure \ref{eprfttiingerror}, and the slightly lower value for Pc:PIC reflecting its broader and more complex spectral profile.

\clearpage
\subsection{EPR Rabi Oscillations}

\begin{figure}[H]
    \centering
    \includegraphics[width=0.6\linewidth]{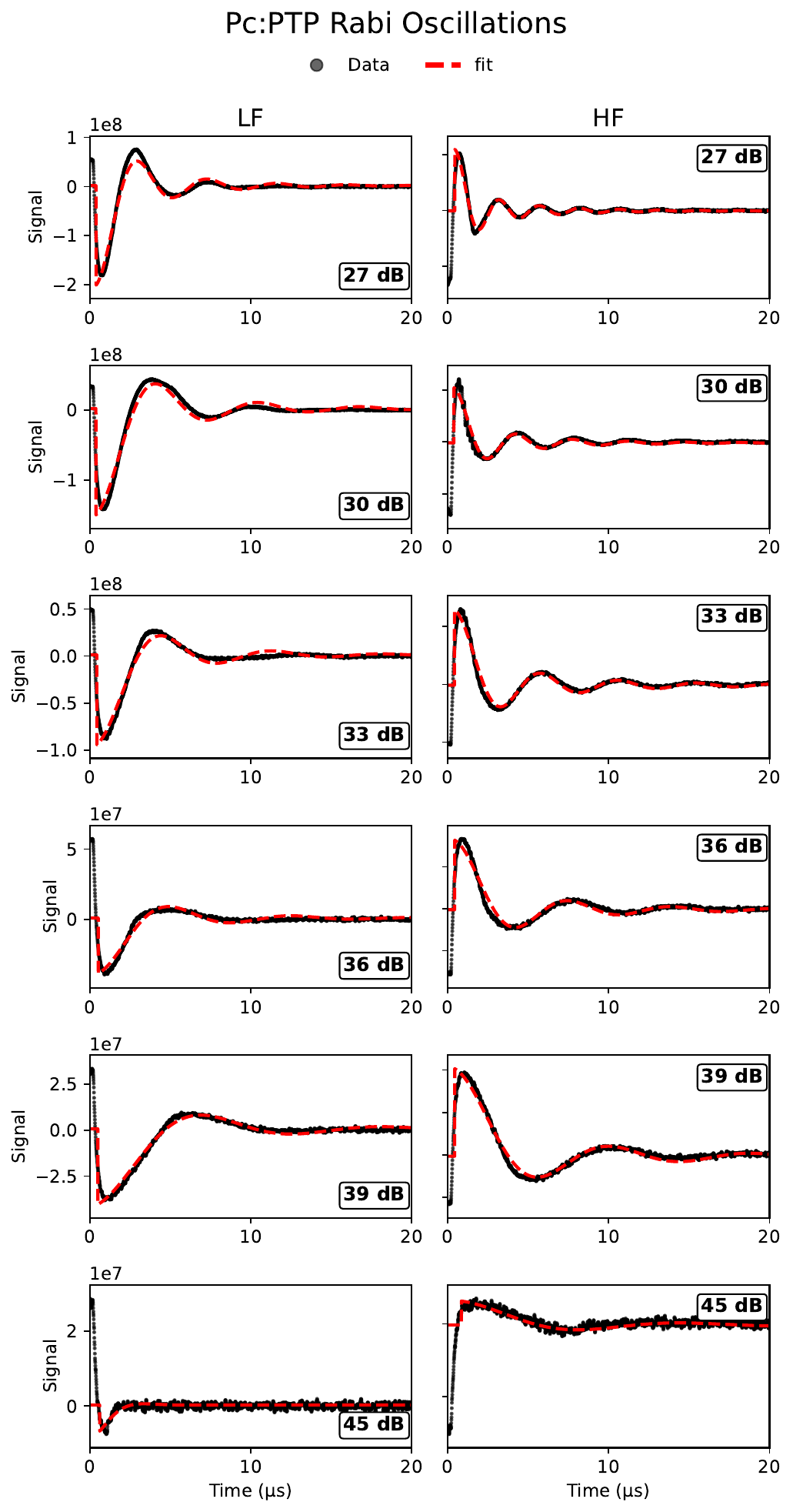}
    \caption{Experimental (black dots) and fitted (red lines) Rabi oscillations observed in the polarisation decay of Pc:PTP by time-resolved EPR spectroscopy under different microwave attenuations.}
    \label{Pc_PTP_RO}
\end{figure}

\begin{figure}[H]
    \centering
    \includegraphics[width=0.6\linewidth]{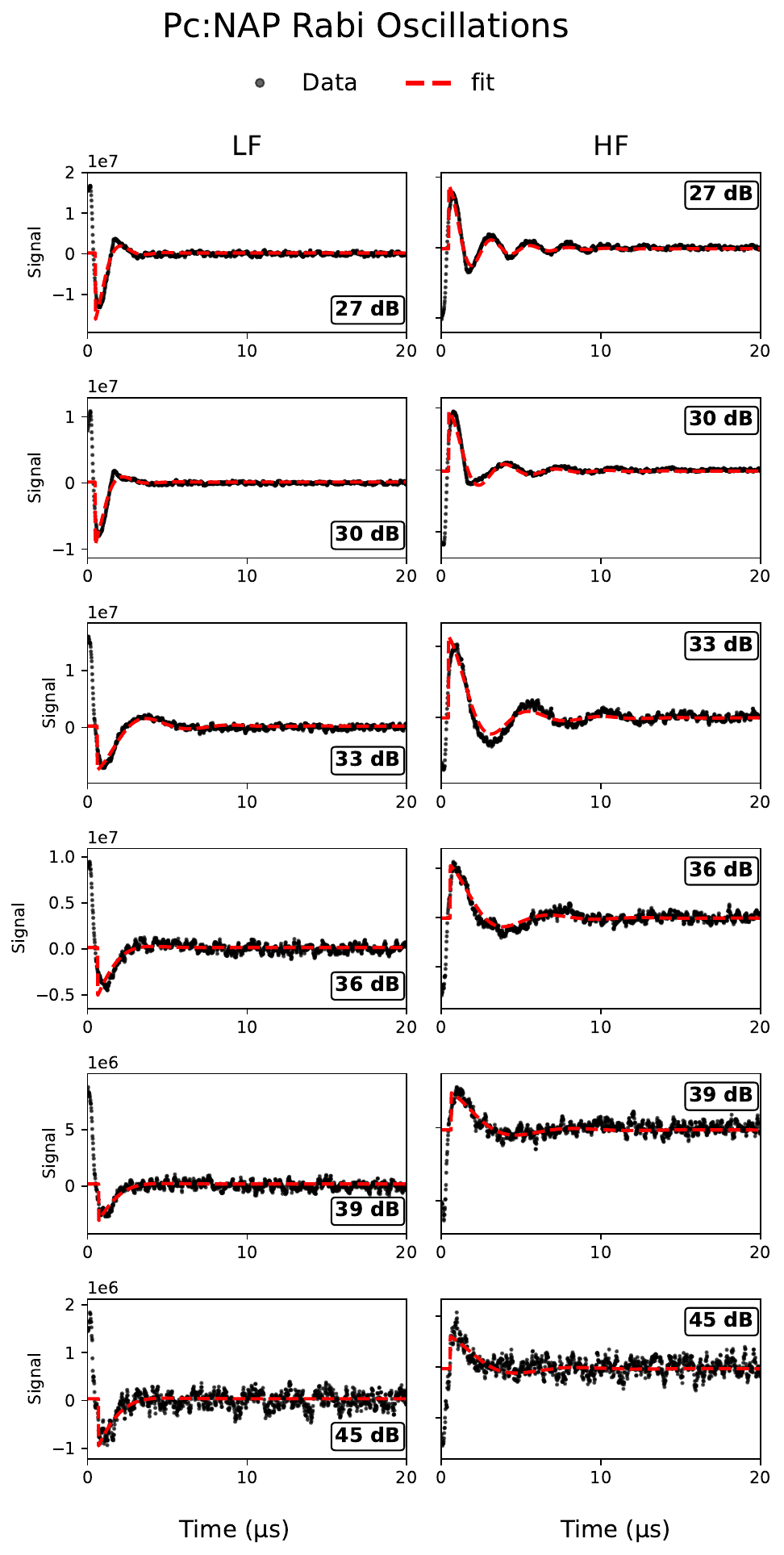}
    \caption{Experimental (black dots) and fitted (red lines) Rabi oscillations observed in the polarisation decay of Pc:NAP by time-resolved EPR spectroscopy under different microwave attenuations.}
    \label{Pc_NAP_RO}
\end{figure}

\begin{figure}[H]
    \centering
    \includegraphics[width=0.6\linewidth]{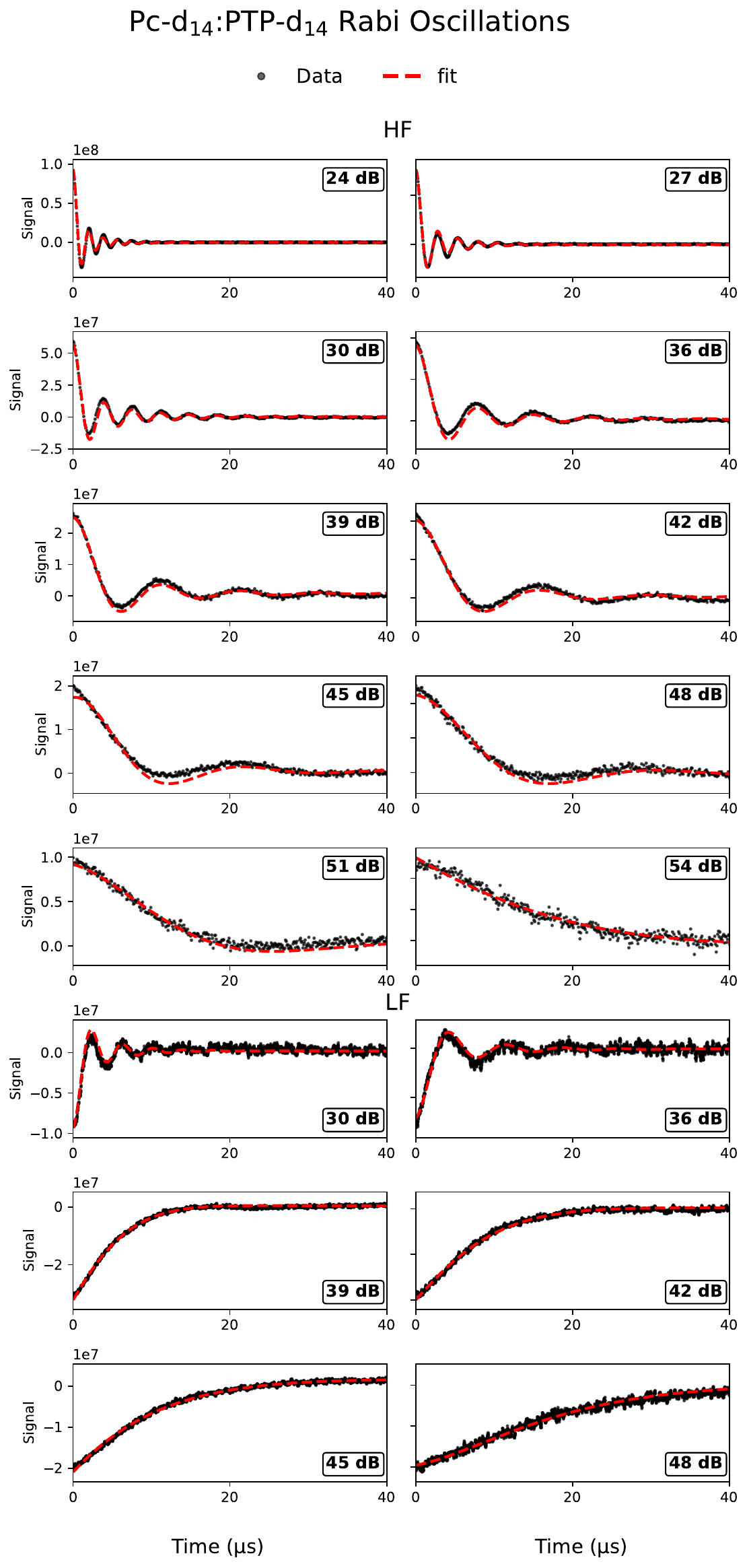}
    \caption{Experimental (black dots) and fitted (red lines) Rabi oscillations observed in the polarisation decay of Pc-d$_{14}$:PTP-d$_{14}$ by time-resolved EPR spectroscopy under different microwave attenuations.}
    \label{Pcd14_PTPd14_RO}
\end{figure}

\begin{figure}[H]
    \centering
    \includegraphics[width=0.7\linewidth]{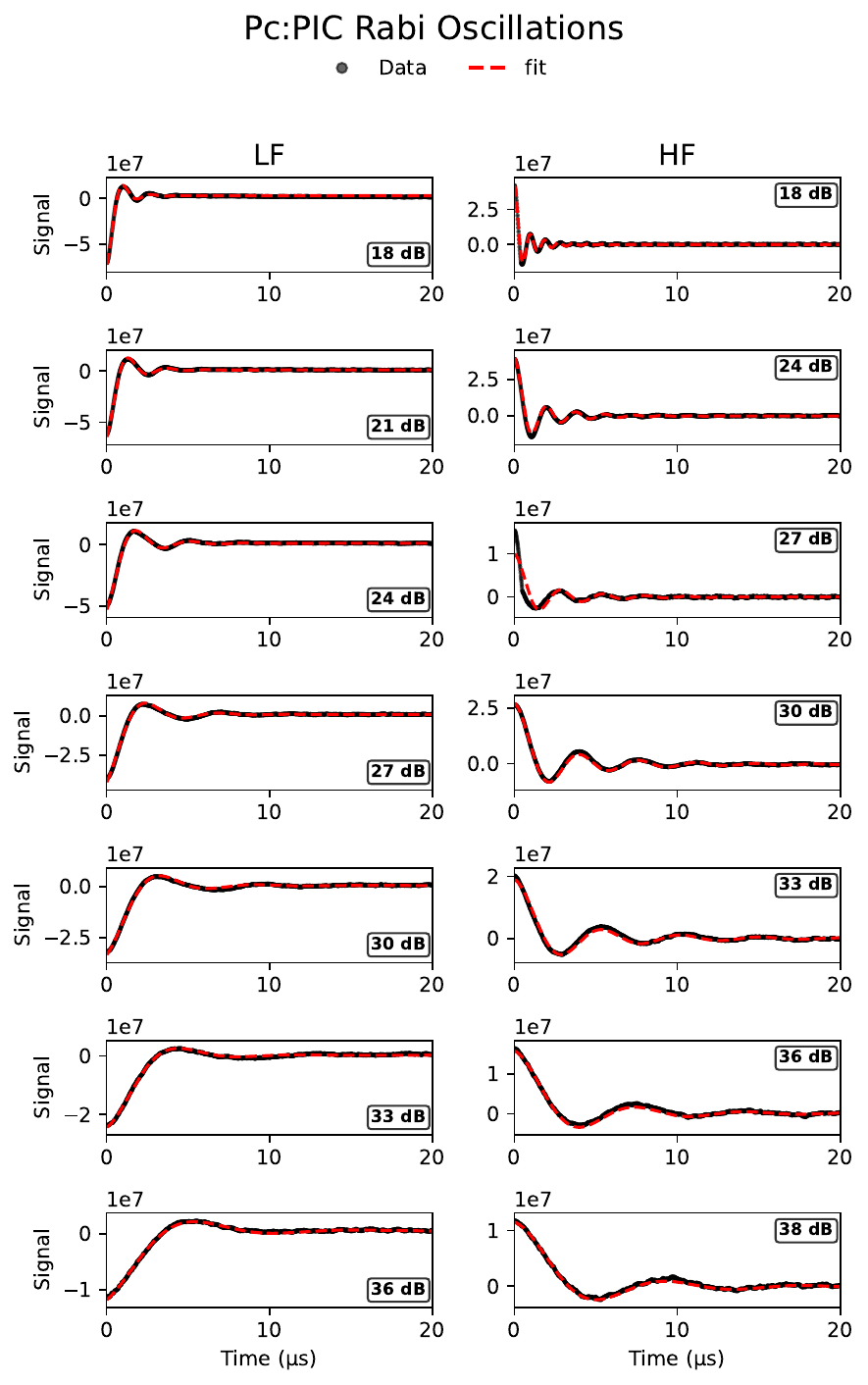}
    \caption{Experimental (black dots) and fitted (red lines) Rabi oscillations observed in the polarisation decay of Pc:PIC measured by time-resolved EPR spectroscopy under different microwave attenuations.}
    \label{Pc_PIC_RO}
\end{figure}

\begin{figure}[H]
    \centering
    \includegraphics[width=0.9\linewidth]{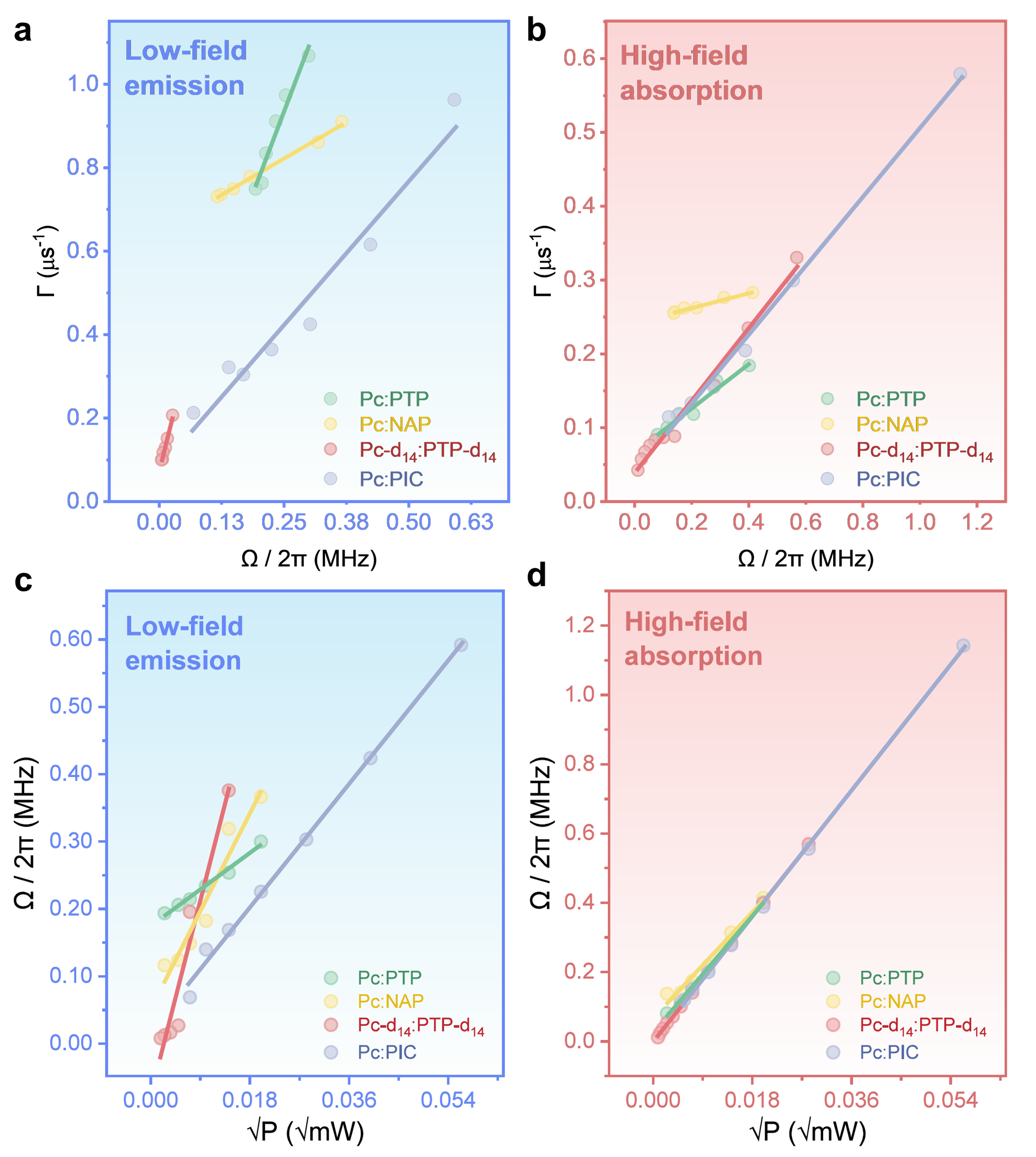}
    \caption{\textbf{Microwave-power-dependent EPR Rabi measurements.} (a, b) Linear dependence of oscillation frequency with applied microwave power, fitted using the equation  $\Gamma$ = $\Delta$$\Omega$/(2$\pi$) + 1/(2$T_{2,Rabi}$) to extract $\Gamma$ and $\Omega$ from microwave-power-dependent oscillations at LF emission (blue) and HF absorption (red) region, where $\Delta$ is the gradient of the fitted linear lines. (c, d) Rabi frequency $\Omega$ as a function of the square root of the applied microwave power.}
    \label{Rabi_and_power_fittings}
\end{figure}

While performing trEPR experiments on each of the materials, Rabi oscillations were observed in the polarisation decay and used to estimate $T_2$. These oscillations were measured under various microwave powers and the most intense low-field (LF) and high-field (HF) features were individually fitted to a zero-order Bessel function to capture the spin-polarisation decay as well as the oscillation frequency, $\Omega$, and the dampening rate, $\Gamma$ (Figures \ref{Pc_PTP_RO}-\ref{Pc_PIC_RO}). The $T_2$ was then extracted from the intercept of the linear relation (Figure \ref{Rabi_and_power_fittings}a-b), previously used by Wang and colleagues \cite{wang2024tailoring}. We found the $T_2$ of the HF absorption feature to be approximately twice that obtained from the LF emission feature. This field dependence is consistent with previous electron-spin-echo studies of Pc:PTP triplet states \cite{kouskov1995pulsed}, where LF transitions were found to be more strongly affected by hyperfine-mediated mixing: the electron spin experiences a less well-defined local magnetic environment due to stronger coupling to surrounding nuclear spins. This reduces the efficiency of the microwave pulse in driving a coherent spin rotation and leads to faster damping of the Rabi oscillations.

Similarly, at both LF and HF regions, the Rabi frequency $\Omega$ increases linearly with the square root of the applied microwave power (Figure \ref{Rabi_and_power_fittings}c-d). The slope of each fitted line reflects the microwave-to-spin conversion efficiency \cite{hyde1989multipurpose}, revealing that all host-guest systems are strongly coupled to microwave field at HF, while they show slightly reduced slopes at LF due to differences in weaker transition dipole coupling. All of these above quantities determine the achievable population inversion and transition linewidth, which ultimately govern the maser gain. 
Under an applied magnetic field, we exploit the fact that the Rabi frequency $\Omega$ and the damping rate $\Gamma$ of the nutation oscillations arising from spin dephasing processes. Accordingly, the samples were scanned at various microwave powers, and all time-domain signal intensities \textit{I(t)} were then fitted to a zero-order Bessel function in the form of:
\begin{equation}
    I(t) = A \cdot J_0[\Omega (t - t_0)] \cdot \exp[-\Gamma (t - t_0)] + C
\label{BESSEL}
\end{equation}
where $A$ denotes the oscillation amplitude determined by the initial spin polarisation, $t_0$ corrects for timing offsets between the microwave pulse and signal acquisition, and $C$ represents a constant background. 

The field dependence of the Rabi damping can be rationalised by considering the hyperfine contribution to the effective triplet spin Hamiltonian. Hyperfine interactions may be separated into secular terms, which primarily shift resonance positions, and non-secular terms, which can mix electron and nuclear spin states \cite{kouskov1995pulsed}. At high magnetic field, the electronic Zeeman interaction defines a robust quantisation axis, so the secular hyperfine terms dominate and non-secular mixing is largely suppressed. As a result, the microwave field drives more coherent spin rotations, leading to stronger Rabi oscillations and slower damping. At lower magnetic field, the Zeeman interaction is less dominant relative to the ZFS and hyperfine interactions. Non-secular hyperfine terms therefore induce stronger electron–nuclear spin mixing, producing a broader distribution of effective local fields and nutation frequencies across the ensemble. Consequently, the LF transition exhibits faster Rabi damping than the HF transition. 

\begin{table}[H]
\centering
\caption{Linear-fit parameters derived from equation 2 for $\Gamma$ versus $\Omega/2\pi$ from X-band transient EPR at LF emission and HF absorption regions.}
\label{tab:Gamma_vs_Omega_fit}
\sisetup{
  detect-weight = true,
  detect-inline-weight = math,
  table-number-alignment = center,
  table-format = 1.5,
}
\setlength{\tabcolsep}{5.5pt}
\renewcommand{\arraystretch}{1.15}

\begin{tabular}{ll
                S[table-format=1.0]  
                S[table-format=1.0]  
                S[table-format=1.0]  
                S[table-format=1.0]} 
\toprule
 &  & \multicolumn{1}{c}{Pc:PTP} & \multicolumn{1}{c}{Pc:NAP} &
      \multicolumn{1}{c}{Pc-d$_{14}$:PTP-d$_{14}$} & \multicolumn{1}{c}{Pc:PIC} \\
\midrule

\multicolumn{6}{l}{\textbf{LF}}\\
\addlinespace[2pt]

\multirow{2}{*}{Intercept} & Value          & 0.147 & 0.648 & 0.072 & 0.078 \\
                           & Standard error & 0.078 & 0.006 & 0.004 & 0.042 \\
\addlinespace[2pt]

\multirow{2}{*}{Slope}     & Value          & 3.15 & 0.696 & 4.82 & 1.38 \\
                           & Standard error & 0.328 & 0.025 & 0.249 & 0.130 \\
\addlinespace[2pt]

\multicolumn{2}{l}{Adj.\ $R^{2}$}        & 0.948 & 0.993 & 0.987 & 0.949 \\
\midrule

\multicolumn{6}{l}{\textbf{HF}}\\
\addlinespace[2pt]

\multirow{2}{*}{Intercept} & Value          & 0.066 & 0.242 & 0.039 & 0.039 \\
                           & Standard error & 0.007 & 0.002 & 0.006 & 0.008 \\
\addlinespace[2pt]

\multirow{2}{*}{Slope}     & Value          & 0.300 & 0.100 & 0.490 & 0.467 \\
                           & Standard error & 0.031 & 0.008 & 0.022 & 0.016 \\
\addlinespace[2pt]

\multicolumn{2}{l}{Adj.\ $R^{2}$}        & 0.949 & 0.966 & 0.982 & 0.993 \\
\bottomrule
\end{tabular}
\end{table}

\clearpage

\newpage
\section{Room-temperature ODMR}

\begin{figure*}[htb!]
    \centering
    \includegraphics{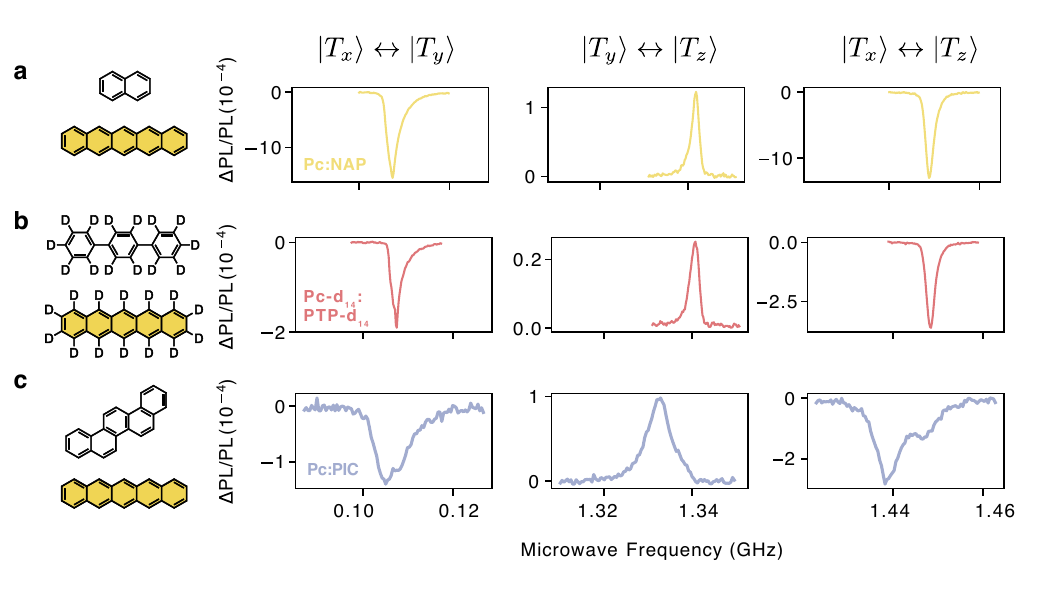}
    \caption{Room-temperature CW-ODMR measurements for \textbf{a} Pc:NAP, \textbf{b}  Pc-d\textsubscript{14}:PTP-d\textsubscript{14} and \textbf{c} Pc:PIC single crystals for all transitions.}\label{CW-ODMR}
\end{figure*}

\begin{figure*}[htb!]
    \centering
    \includegraphics[width=\columnwidth]{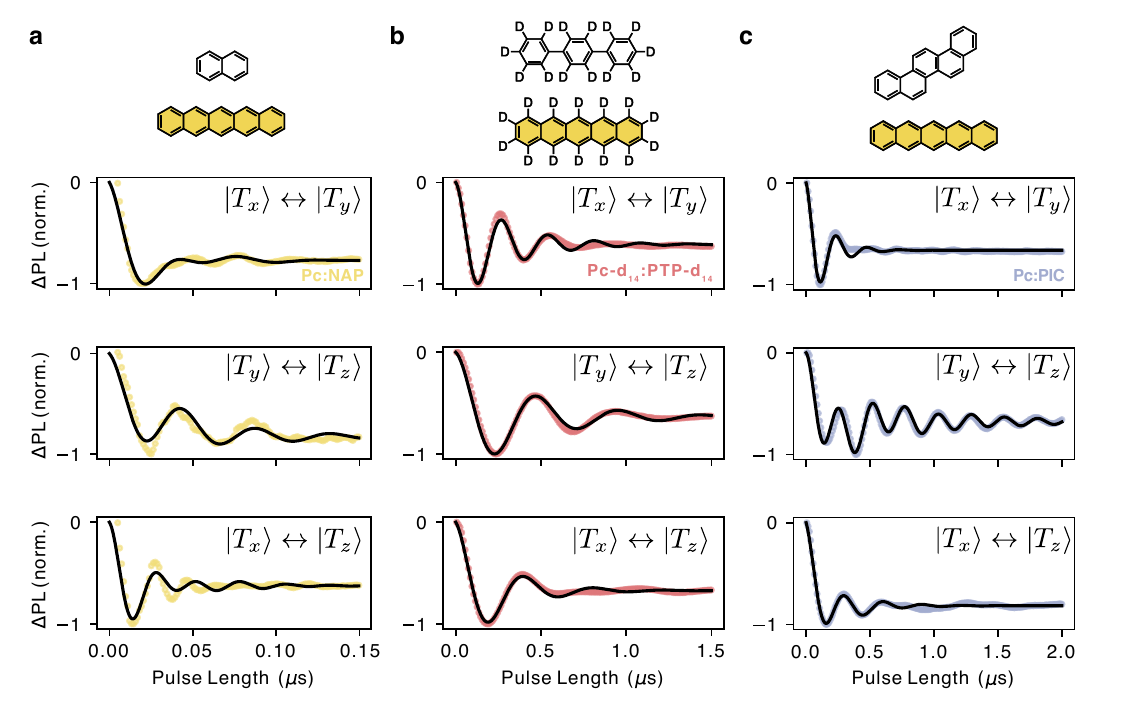}
    \caption{\textbf{Room-temperature optically detected Rabi oscillations.} Rabi oscillations for \textbf{a} Pc:NAP, \textbf{b} Pc-d\textsubscript{14}:PTP-d\textsubscript{14} and \textbf{c} Pc:PIC single crystals for all transitions. Black lines are fits to damped cosine functions with a single component, $A \exp\left(\frac{-t}{T_2^{\text{MW}}}\right)\cos(\omega_Rt)+B$, or two components, $A \left[\exp\left(\frac{-t}{T_2^{\text{MW,1}}}\right)\cos(\omega_{R,1}t) + \exp\left(\frac{-t}{T_2^{\text{MW,2}}}\right)\cos(\omega_{R,2}t)\right]+B$ where $t$ is the microwave pulse length, and the amplitude, $A$, offset, $B$, decay time, $T_2^{\text{MW}}$ and Rabi oscillation frequency, $\omega_R$, were fit variables. The fits were used to determine population inversion following microwave pulses on each transition.}
\label{Rabi}
\end{figure*}

\clearpage
\subsection{Optically Detected Relaxation Measurements}

\begin{figure*}[h]
    \centering
    \includegraphics{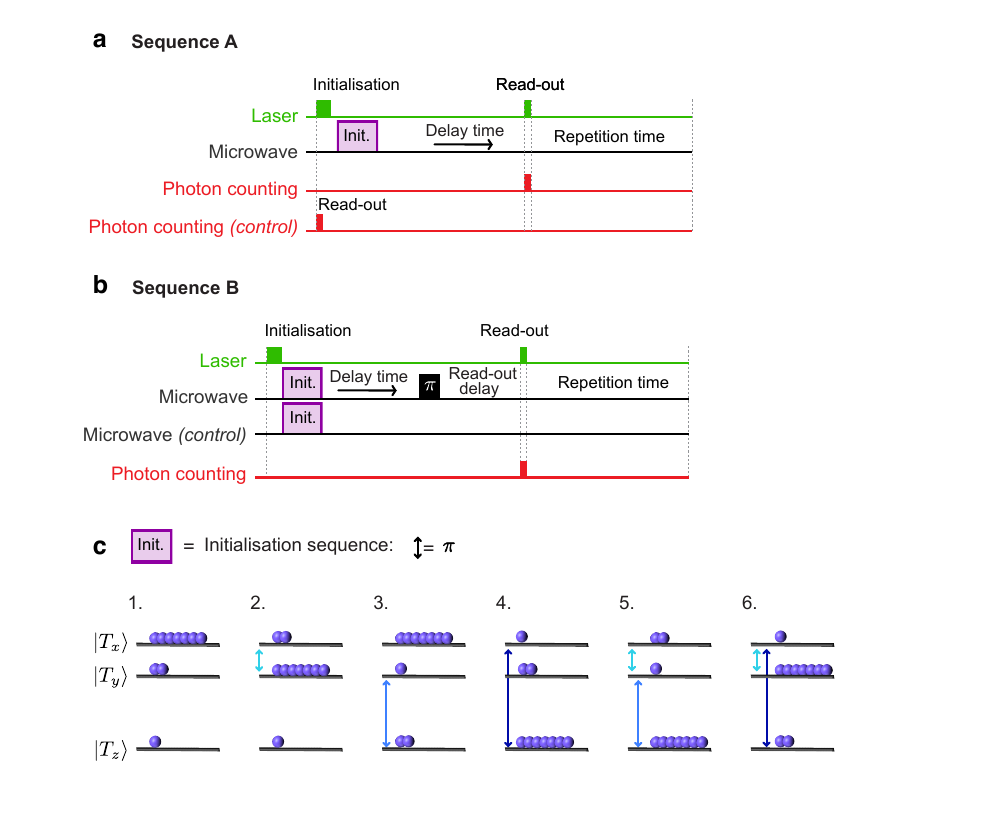}
    \caption{\textbf{ODMR pulse sequences for quantifying triplet spin dynamics \cite{mann2025chemically}.} \textbf{a, b} Pulse sequences  A and B used to determine the dynamics. \textbf{c} Initialisation sequences 1-6, used to prepare the triplet sublevels in six different states (illustrated using purple circles to represent the relative sublevel populations).}\label{pulse-sequences}
\end{figure*}

\newpage
\begin{figure*}[htb!]
    \centering
    \includegraphics{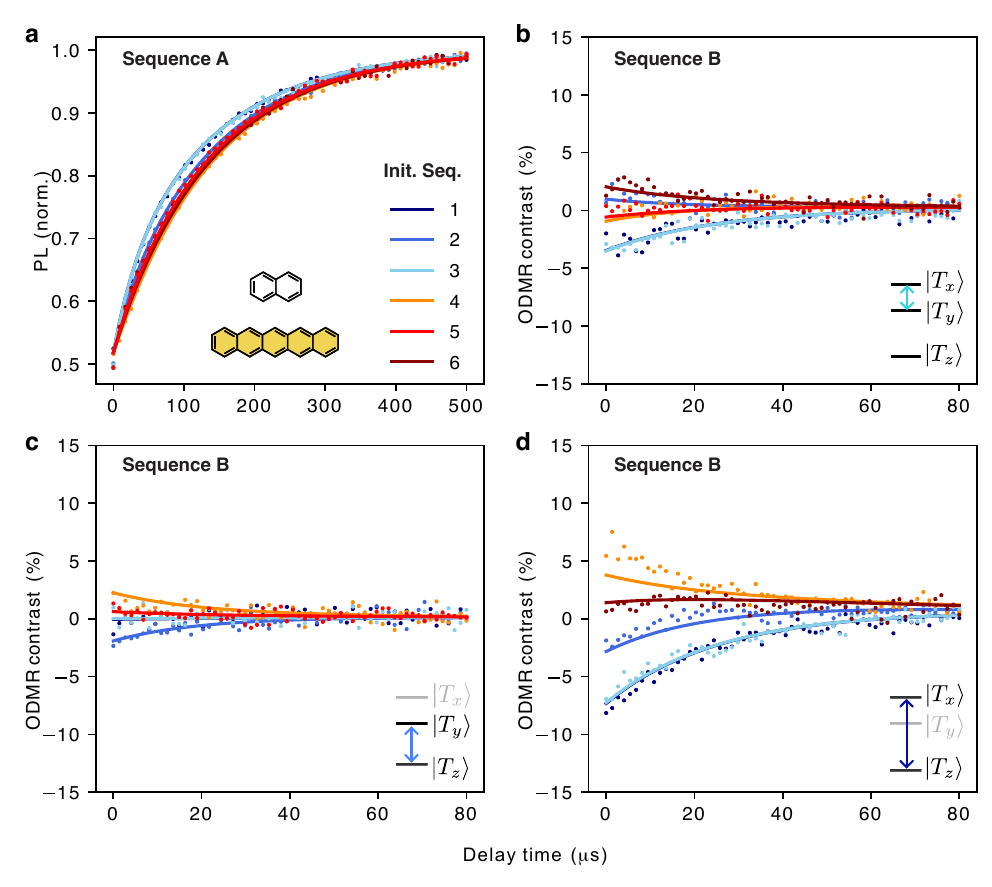}
    \caption{\textbf{Room temperature optically detected relaxation measurements for Pc:NAP.} Relaxation curves recorded using \textbf{a} Sequence A, and  \textbf{b, c, d} Sequence B with ODMR measurement on the \textbf{b} $|T_x\rangle \leftrightarrow |T_y\rangle$, \textbf{c} $|T_y\rangle \leftrightarrow |T_z\rangle$ and \textbf{d} $|T_x\rangle \leftrightarrow |T_z\rangle$ transition. Fits are shown by solid lines.}\label{NAP_relaxation}
\end{figure*}

\begin{figure*}[htb!]
    \centering
    \includegraphics{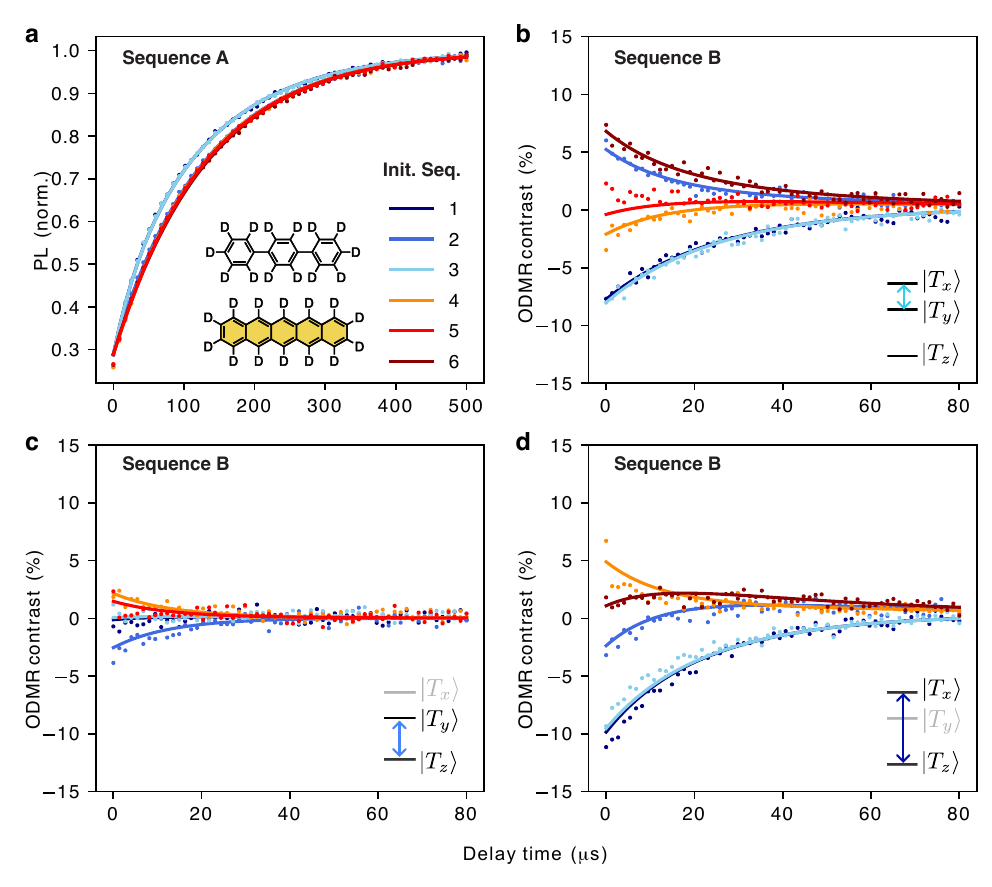}
    \caption{\textbf{Room temperature optically detected relaxation measurements for Pc-d\textsubscript{14}:PTP-d\textsubscript{14}.} Relaxation curves recorded using \textbf{a} Sequence A, and  \textbf{b, c, d} Sequence B with ODMR measurement on the \textbf{b} $|T_x\rangle \leftrightarrow |T_y\rangle$, \textbf{c} $|T_y\rangle \leftrightarrow |T_z\rangle$ and \textbf{d} $|T_x\rangle \leftrightarrow |T_z\rangle$ transition. Fits are shown by solid lines.}\label{d14_relaxation}
\end{figure*}

\newpage
\begin{figure*}[htb!]
    \centering
    \includegraphics{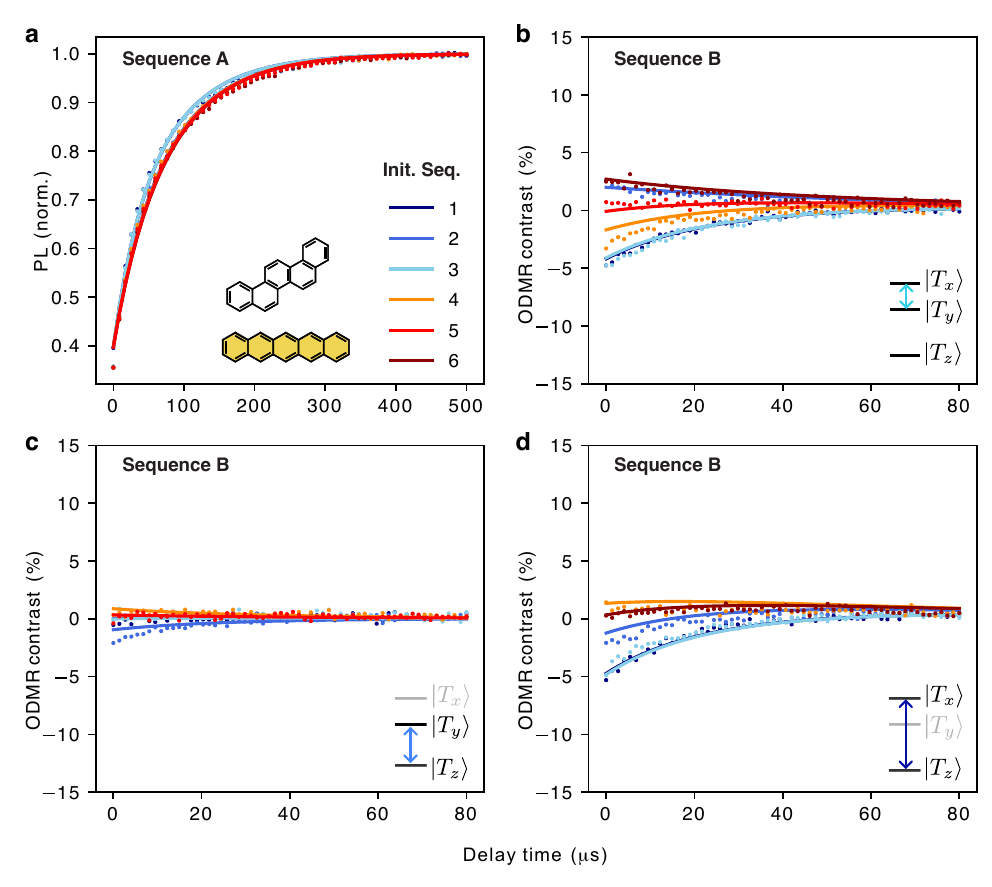}
    \caption{\textbf{Room temperature optically detected relaxation measurements for Pc:PIC.} Relaxation curves recorded using \textbf{a} Sequence A, and  \textbf{b, c, d} Sequence B with ODMR measurement on the \textbf{b} $|T_x\rangle \leftrightarrow |T_y\rangle$, \textbf{c} $|T_y\rangle \leftrightarrow |T_z\rangle$ and \textbf{d} $|T_x\rangle \leftrightarrow |T_z\rangle$ transition. Fits are shown by solid lines.}\label{picene_relaxation}
\end{figure*}

\begin{table*}[h]
\caption{\textbf{Room-temperature zero-field spin dynamics of Pc:PTP, Pc-d\textsubscript{14}:PTP-d\textsubscript{14}, Pc:NAP and Pc:PIC.} Extracted populations of the singlet ground state ($P_0$) and triplet sublevels ($P_x, P_y, P_z$) following photoexcitation.}
\begin{ruledtabular}
\begin{tabular}{ccccc}
& $|S_0 \rangle$ Population & \multicolumn{3}{c}{Normalized triplet populations\textsuperscript{a}} \\
&$P_0$ & $P_x$ & $P_y$ & $P_z$  \\ 
\hline
Pc:PTP \cite{mann2025chemically}               &$0.347 \pm 0.001$ & $0.732 \pm 0.003$ & $0.159 \pm 0.002$ & $0.109 \pm 0.002$  \\
Pc:NAP                                                 &$0.518 \pm 0.001$ & $0.706 \pm 0.009$ & $0.151 \pm 0.007$ & $0.143 \pm 0.007$  \\
Pc-d\textsubscript{14}:PTP-d\textsubscript{14}         &$0.289 \pm 0.001$ & $0.720 \pm 0.008$ & $0.152 \pm 0.005$ & $0.127 \pm 0.006$  \\
Pc:PIC                                                 &$0.397 \pm 0.001$ & $0.556 \pm 0.005$ & $0.217 \pm 0.005$ & $0.227 \pm 0.004$  \\
\end{tabular}
\end{ruledtabular}
\label{table2}
\footnotetext{The triplet sublevel populations are normalized such that $P_x+P_y+P_z=1$}
\label{table2}
\end{table*}

\clearpage
\subsection{CW-ODMR linewidth and $T_2$ analysis}

\begin{figure}[H]
    \centering
    \includegraphics{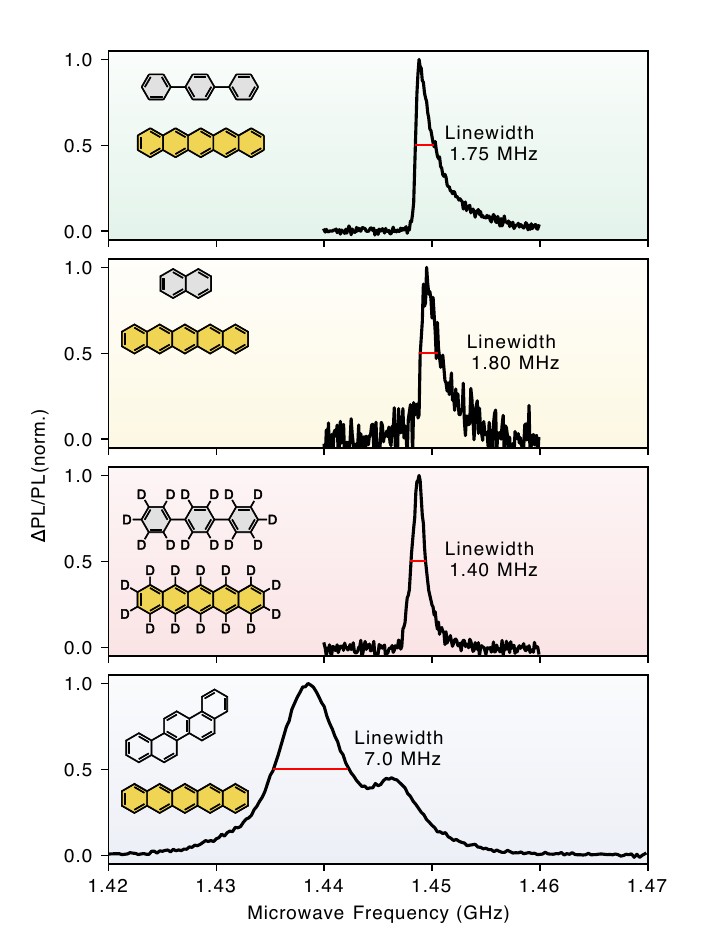}
    \caption{\textbf{Room-temperature zero-field CW-ODMR linewidth analysis for the $\ket{T_x} \leftrightarrow \ket{T_z}$ transition in Pc:PTP, Pc:NAP, Pc-d\textsubscript{14}:PTP-d\textsubscript{14}, and Pc:PIC crystals.}  The reported linewidths correspond to the narrowest values obtained after checking for and minimising laser- and microwave-induced power broadening.
    }
    \label{linewidth}
\end{figure}
\clearpage
\newpage

\begin{figure}  
  \centering
  \includegraphics[width=0.8\textwidth]{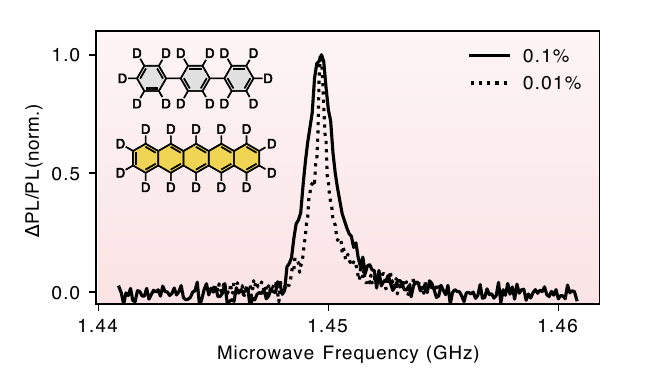}
    \caption{\textbf{Guest-concentration dependence of the room-temperature zero-field CW-ODMR linewidth in Pc-d$_{14}$:PTP-d$_{14}$.} Normalised ODMR spectra are shown for guest concentrations of 0.1\% and 0.01\%. The corresponding linewidths extracted from the spectra are 1.4\,MHz and 0.55\,MHz, respectively. The reported linewidths correspond to the narrowest values obtained after checking for laser- and microwave-induced power broadening.}
  \label{ODMR_conc}
\end{figure}

\begin{table}[h]
\caption{CW-ODMR linewidths for various pentacene-doped samples as shown in Figure\,\ref{linewidth} and \ref{ODMR_conc}.}
\label{tab:linewidths}
\begin{ruledtabular}
\begin{tabular}{ll}
Sample & Linewidth (MHz) \\
\hline
0.1\% Pc:PTP & 1.75 (1.9 \cite{singh2025pentacene}) \\
0.04\% Pc:NAP & 1.8 \\
0.004\% Pc-d$_{14}$:NAP & (0.6 \cite{li_robust_2026}) \\
0.1\% Pc-d$_{14}$:PTP-d$_{14}$ & 1.4\\
0.01\% Pc-d$_{14}$:PTP-d$_{14}$ & 0.55 \\
0.1\% Pc:PIC & 7.0 \\
\end{tabular}
\end{ruledtabular}
\end{table}

\clearpage
\subsection{Optically Detected Hahn-Echo Decay}

\begin{figure*}[htb!]
    \centering
    \includegraphics{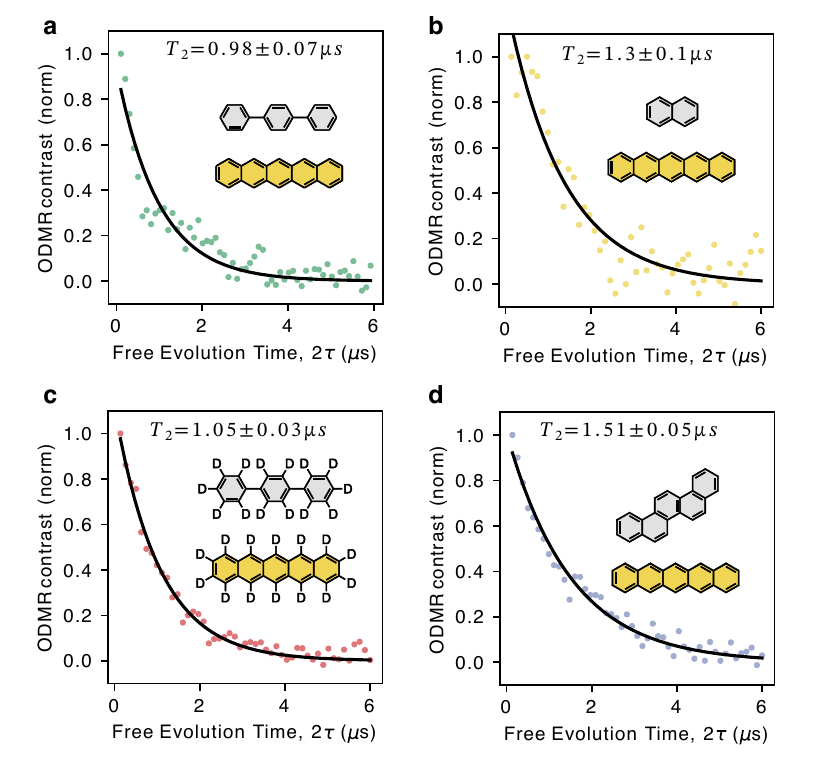}
    \caption{Room-temperature zero-field optically detected Hahn echos for \textbf{a} Pc:PTP, \textbf{b} Pc:NAP, \textbf{c} Pc-d\textsubscript{14}:PTP-d\textsubscript{14} and \textbf{d} Pc:PIC.}
    \label{T2s}
\end{figure*}

\clearpage
\subsection{EPR and ODMR Comparison}

\begin{figure}[H]
    \centering
    \includegraphics[width=0.9\linewidth]{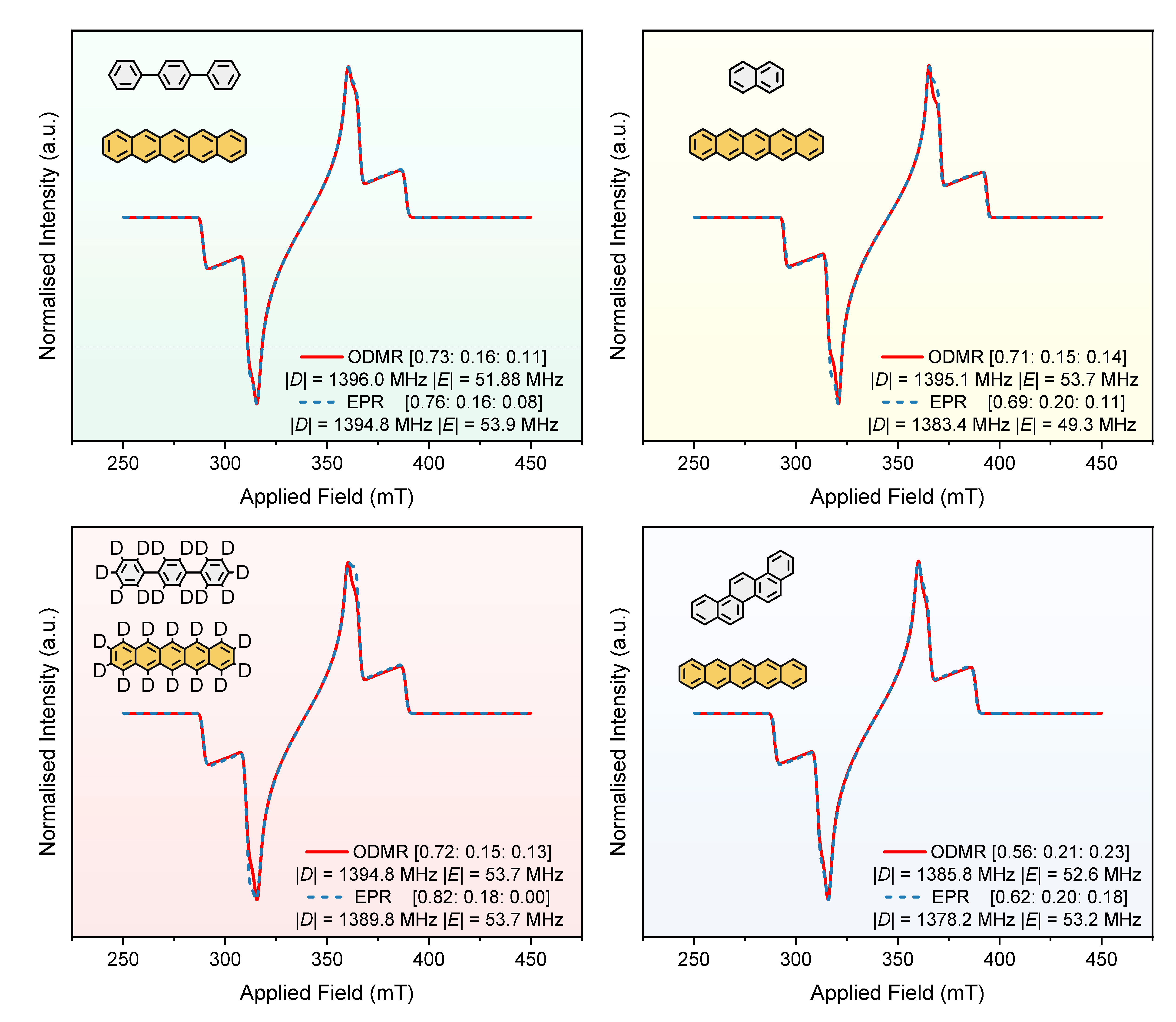}
    \caption{\textbf{Simulations of powder X-band EPR spectrum for all host-guest systems with different ZFS parameters and triplet population ratios.} Simulations were carried out in \textit{EasySpin} \cite{stoll2006easyspin} at 9.5 GHz. An isotropic \textit{g} value of 2.0023 was used, with a Lorentzian broadening parameter of 2 MHz to reproduce the linewidth.}
    \label{EPR_vs_ODMR}
\end{figure}

\clearpage
\section{Population Kinetics and Maser Cooperativity}

\begin{figure}[ht!]
    \centering
    \includegraphics[width=0.85\linewidth]{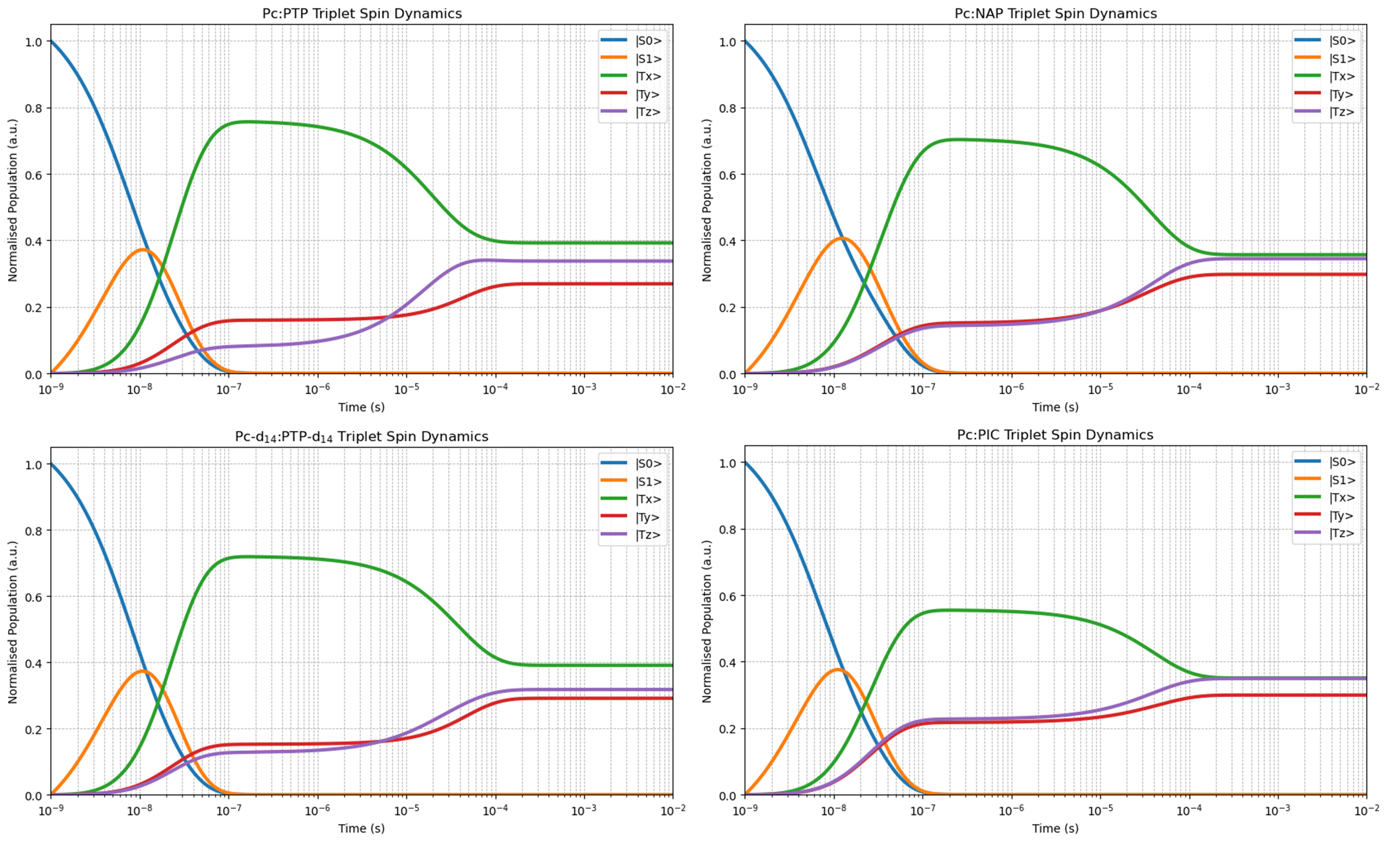}
    \caption{\textbf{Population kinetics for all host-guest systems under continuous light excitation.} The initial triplet-sublevel populations were taken from the ODMR-derived populations, while the experimentally determined rates $k_i$ and $w_{ij}$ are reported in Table II in the main text. For Pc:NAP, $k_{\mathrm{ISC}}$ was estimated using the ISC rate reported by Takeda \textit{et al.} \cite{takeda2002zero}. The $k_{\mathrm{ISC}}$ for Pc:PIC was assumed to be the same as that of Pc:PTP. A pumped rate of $k_{01}=1.1\times10^{8}\ \mathrm{s}^{-1}$ was used throughout.}
    \label{dynamics_sim}
\end{figure}

The population dynamics of the singlet and triplet manifolds in Figure \ref{dynamics_sim} can be described by the following coupled rate equations:
\begin{equation}
\begin{aligned}
\frac{dS_0}{dt} &= -k_{01} S_0 + k_{10} S_1 + k_x T_x + k_y T_y + k_z T_z \\
\\
\frac{dS_1}{dt} &= k_{01} S_0 - (k_{10} + k_{\mathrm{ISC}}) S_1 \\
\\
\frac{dT_x}{dt} &= k_{\mathrm{ISC}} P_x S_1 
- (k_x + \omega_{xy} + \omega_{xz}) T_x
+ \omega_{yx} T_y + \omega_{zx} T_z \\
\\
\frac{dT_y}{dt} &= k_{\mathrm{ISC}} P_y S_1
- (k_y + \omega_{yx} + \omega_{yz}) T_y
+ \omega_{xy} T_x + \omega_{zy} T_z \\
\\
\frac{dT_z}{dt} &= k_{\mathrm{ISC}} P_z S_1
- (k_z + \omega_{zx} + \omega_{zy}) T_z
+ \omega_{xz} T_x + \omega_{yz} T_y
\end{aligned}
\label{rate_equations}
\end{equation}

\clearpage
The spin population inversion $\Delta N(t)$ for the $\ket{T_x}\leftrightarrow\ket{T_z}$ transition was obtained by solving Equation \ref{rate_equations}, and the final maser cooperativity $\eta_{\mathrm{maser}}$ was calculated according to Equation 1 of the "Toward Continuous Operation" subsection in the main text, and using parameters listed in Table \ref{tab:coop_params}.

\begin{table}[h]
\caption{Parameters used for the simulation of the maser cooperativity $\eta_{\mathrm{maser}}$ in Figure 7.}
\label{tab:coop_params}
\begin{ruledtabular}
\begin{tabular}{lll}
Parameter & Symbol & Value / Expression \\
\hline
Quality factor & $Q$ & 3500 \\
Magnetic filling factor & $\eta_{\mathrm{mag}}$ & 0.25 \\
Magnetic mode volume & $V_{\mathrm{mag}}$ & $2.5 \times 10^{-7}$ m$^{3}$\\
$\ket{T_x}$ $\leftrightarrow$ $\ket{T_z}$ transition linewidth (taken \\
from FWHM in \ref{linewidth}) & $\Delta f$ & 
Pc:PTP =  1.75 MHz\\
& & Pc:NAP = 1.8 MHz \\
& & Pc-d$_{14}$:PTP-d$_{14}$ = 1.4 MHz \\
& & Pc:PIC = 7 MHz \\
Time-dependent normalised spin inversion & $\Delta N(t)$ & 
$\rho_{\mathrm{spin}} n_{\mathrm{mol}} V_{\mathrm{mag}}[P_x(t)-P_z(t)]$ \\
\end{tabular}
\end{ruledtabular}
\end{table}
\clearpage


\newpage
\section{Maser Experiment}
\begin{figure}[hb!]
    \centering
    \includegraphics[width=0.65\linewidth]{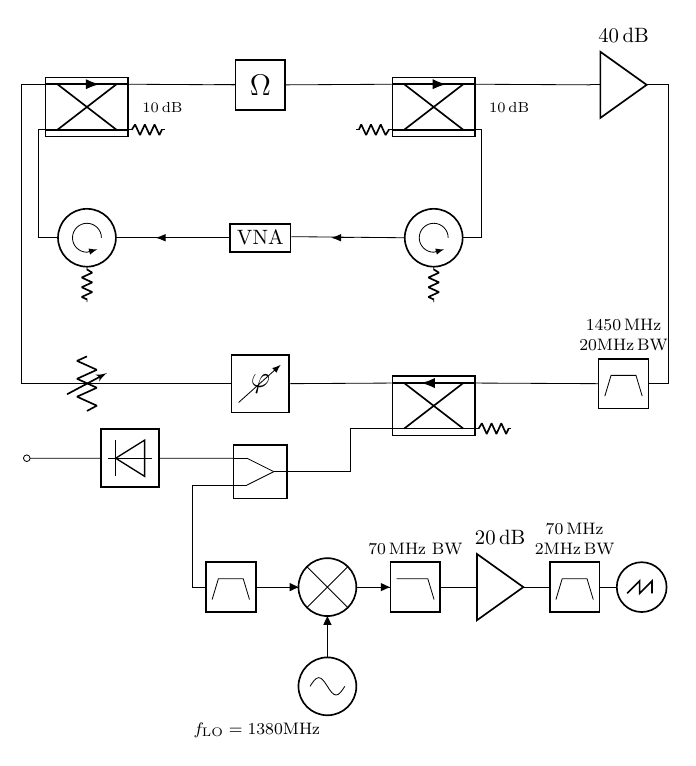}
    \caption{Set up for \textit{Q}-boosted maser experiments with a superheterodyne receiver.}
    \label{q_boost}
\end{figure}

The maser threshold and maximum output power experiments were performed using the “maser-in-a-shoebox” set-up \cite{ng2024maser}. All host–guest systems investigated here adopt a monoclinic lattice upon incorporation of pentacene, in space group P2$_1$/a \cite{moro2022room}. In each case, the dopant molecules occupy substitutions in well-defined orientations relative to their host framework, such that the molecular $y$-axis of at least one site of pentacene remains suitably aligned to couple with the axisymmetric transverse electric (TE$_{01\delta}$) mode of a high-\textit{Q} strontium titanate (STO) resonator \cite{breeze2015enhanced}, yielding a non-zero matrix element $\langle T_x \lvert \hat{S}_y \rvert T_z \rangle \neq0$, therefore allowing $\ket{T_x}$ $\leftrightarrow$ $\ket{T_z}$ transition. A thin coating of photochemically resistant optical coupling fluid (\textit{Santovac 5} polyphenyl ether diffusion-pump oil) was used to fill the gap between crystals and STO resonator. The set-up of \textit{Q}-boosting experiment is shown in Figure \ref{q_boost}.

\clearpage
\bibliography{References} 
\bibliographystyle{rsc}